\documentclass[%
 preprint, 
 showkeys,
 superscriptaddress,
 amsmath,amssymb,
 aps, physrev,
]{revtex4-2}

\usepackage{graphicx}
\usepackage{dcolumn}
\usepackage{bm}
\usepackage{hyperref}
\usepackage[mathlines]{lineno}
\usepackage{listings}
\usepackage{siunitx}

\usepackage{underscore}
\usepackage[english]{babel}
\usepackage[ruled,lined]{algorithm2e}

\begin{document}

\title{\textbf{Exponential capacity scaling of classical GANs compared to hybrid latent style-based quantum GANs}
}

\author{Milan Liepelt}
 \email{milan.liepelt@artidis.com}
\affiliation{ARTIDIS, Hochbergerstrasse 60c, CH-4057 Basel, Switzerland}
\affiliation{Center for Quantum Computing and Quantum Coherence (QC2),\\
University of Basel, Klingelbergstrasse 82, CH-4056 Basel, Switzerland}%
\author{Julien Baglio}%
 \email{julien.baglio@unibas.ch}
\affiliation{Center for Quantum Computing and Quantum Coherence (QC2),\\
University of Basel, Klingelbergstrasse 82, CH-4056 Basel, Switzerland}%
\affiliation{
QuantumBasel, Schorenweg 44b, CH-4144 Arlesheim, Switzerland}%

\date{\today}

\begin{abstract}
Quantum generative modeling is a very active area of research in looking for practical advantage in data analysis. Quantum generative adversarial networks (QGANs) are leading candidates for quantum generative modeling and have been applied to diverse areas, from high-energy physics to image generation. The latent style-based QGAN, relying on a classical variational autoencoder to encode the input data into a latent space and then using a style-based QGAN for data generation has been proven to be efficient for image generation or drug design, hinting at the use of far less trainable parameters than their classical counterpart to achieve comparable performance, however this advantage has never been systematically studied. We present in this work the first comprehensive experimental analysis of this advantage of QGANS applied to SAT4 image generation, obtaining an exponential advantage in capacity scaling for a quantum generator in the hybrid latent style-based QGAN architecture. Careful tuning of the autoencoder is crucial to obtain stable, reliable results. Once this tuning is performed and defining training optimality as when the training is stable and the FID score is low and stable as well, the optimal capacity (or number of trainable parameters) of the classical discriminator scales exponentially with respect to the capacity of the quantum generator, and the same is true for the capacity of the classical generator. This hints toward a type of quantum advantage for quantum generative modeling.
\end{abstract}


\maketitle
\pagebreak

\section{Introduction}

Generative modeling has become a pillar in modern artificial intelligence (AI) over multiple areas of science and technology and is at the heart of the wide application of large language models~\cite{electronics14183580} or modern image generation in multiple domains including medical data~\cite{karmakar2025role} of for de-novo drug design in generative chemistry~\cite{Tang2024,Gangwal2024}. Generative Adversarial Networks (GANs)~\cite{goodfellow2014generativeadversarialnetworks} are a prominent example of this class of AI models applied to high-quality image generation~\cite{10.1016/j.cag.2023.05.010}, or drug design~\cite{TRIPATHI2022100045,kotkondawar2025generative}.  They are a set of two networks, the generator trained to produce output data from input random noise distribution, and the discriminator classifying the output data from the generator against the input training data of the GAN. The training procedure is an adversarial game where the discriminator has to distinguish between real data (from the input training set) and fake data (output from the generator), while the generator has to fool the discriminator into believing that the generated data is authentic. GANs are powerful but face several challenges, see \cite{TRIPATHI2022100045} for an overview of GANs in the context of drug generation or \cite{Saad2024} for (biomedical) image synthesis: complex training patterns of the adversarial process leading to non-convergence; mode collapse where the generator effectively only generates a sub-sample of the entire diversity of the training set; non-robust training with great sensitivity to the hyperparameters; barren plateaus where vanishing gradients prevent the training procedure to optimally navigates in the landscape of the loss function to find its optimum. Some of these challenges have been addressed by introducing the Wasserstein distance as the backbone of the loss function (WGAN)~\cite{Arjovsky2017}, including also gradient penalty (WGAN-GP)~\cite{gulrajani2017improvedtrainingwassersteingans}. Using the power of latent space, that is projecting the input training dataset in an abstract vector space on which the training is done before decoding the output back to the practical, physical space, also helps addressing model collapse and produces high-quality images, see \cite{Prykhodko2019} for an application in drug design. The style-based architecture~\cite{8977347}, combined with this latent-space approach, has defined a new state of the art for GANs in image generation with styleGAN2~\cite{9156570}, using data re-uploading  where the input noise distribution for the generator is inserted in all layers and not only in the first initial layer of the neural network.

Quantum generative modeling~\cite{amin2018quantum,liu2018differentiable,barthe2025parameterized,demidik2025expressive,demidik2025sample} has emerged as an active area of research to overcome these challenges even further, in particular quantum GANs (QGANs)~\cite{dallaire2018quantum,
lloyd2018quantum,hu2019quantum,zoufal2019quantum,niu2022entangling,chaudhary2023towards}
They are a class of variational quantum algorithms, relying on the rules of quantum mechanics, in particular superposition and entanglement, in which output distributions are
defined by measurement statistics of quantum states. These new architectures open up new opportunities for powerful generative neural networks. Quantum GANs are also promising for generative drug discovery~\cite{Diptanshu_Sikdar,9520764,10556803}. Spectral analysis of the output of neural networks has revealed that quantum machine learning models are able to probe different frequency modes when compared to classical networks, and that this feature seems to be related to the potential for higher expressivity of quantum machine learning~\cite{mhiri2025constrained,jaderberg2024let, xu2024frequency, duffy2025spectral}. A growing body of theoretical and empirical work also suggests that quantum machine learning models can exploit the exponentially large Hilbert space of quantum systems as a high-dimensional feature space, enabling richer data representations than classical models. Schuld \& Killoran (2019) formalized this connection by interpreting quantum state encoding as nonlinear feature maps analogous to kernel methods, which can be classically intractable to compute and facilitate learning with fewer data points~\cite{schuld2019quantumfeature}. Subsequent studies have shown that quantum models can generalize effectively from small training sets, sometimes beyond what classical generalization theory predicts, indicating potential sample-efficiency advantages~\cite{gil2024understanding}. Moreover, analytical approaches based on quantum Fisher information tie circuit structure to generalization behavior, providing insight into how quantum models might require less data to achieve accurate predictions~\cite{Haug2024}. From this perspective, quantum models may
access richer representations than classical models with comparable parameter counts.

The style-based approach has also been introduced for QGANs~\cite{BravoPrieto2022,baglio2024} in the context of data augmentation for high-energy physics and later on combined with latent approach for image generation~\cite{chang2024latentstylebasedquantumgan}. In both cases it has produced results which are on par, if not better, than their classical counterpart, with a small number of trainable parameters. Preliminary results indicate that the latent style-based QGAN can also very efficient for drug design with far less trainable parameters than a classical latent GAN~\cite{baglio2025}. It shall be noted that QGANs can appear in various variants, with either the discriminator being a quantum network, or the generator, or both networks. The hybrid architecture with a classical discriminator and a quantum generator has so far been the most appealing and was adopted in e.g. \cite{BravoPrieto2022,baglio2024,chang2024latentstylebasedquantumgan,baglio2025}. This avoids the need to encode high-dimensional, complex data such as images into quantum states and reduces the complexity in training quantum networks. This asymmetry also substantially reduces the quantum resources required and makes hybrid QGANs suitable candidates for near-term quantum hardware. In all these cases, a reduced number of trainable parameters is quite desirable for a neural network as it increases its explainability. However, to the best of our knowledge, no systematic study of this feature for QGANs has been performed so far.

The goal of our paper is to fill this gap by performing a systematic experimental evaluation of the capacity (the number of trainable parameters) scaling of GANs, using satellite image generation as a use case and the SAT4 open-source dataset~\cite{sat4,liu2020deepsat}. We use an autoencoder to encode the images in the abstract latent vector space, and our main metrics to gauge the quality of the images produced by our (Q)GANs is the Fréchet Inception Distance (FID)~\cite{heusel2017gans}, capturing the similarity between generated images and real ones better than the original Inception score metrics~\cite{NIPS2016_8a3363ab}. We have also used the Jensen–Shannon Divergence (JSD)~\cite{jsd1991} as a secondary metrics.
Expending on \cite{chang2024latentstylebasedquantumgan}, we have systematically analyzed both the autoencoder and the (Q)GAN architectures to find optimal parameters for a stable training and a stable evolution of the FID score. The natural choice for the autoencoder structure is based on convolutional neural networks as we are dealing with images. It has been shown that their quantum counterpart are classically simulable~\cite{bermejo2024quantumconvolutionalneuralnetworks}, motivating our choice to keep the autoencoder as a classical neural network and only use a parameterized quantum circuit for the quantum generator, as done in \cite{chang2024latentstylebasedquantumgan,baglio2025}. We have investigated both classical and quantum generator architectures and for both case we have found that the careful tuning of the autoencoder is crucial to obtain stable results in the training of the (Q)GAN. Once this tuning is performed we define GAN training optimality as when the training is stable and the FID score is low and stable as well. Performing systematic experiments in the ideal noiseless case with a quantum simulator for a latent dimension of 24, we have found that classical discriminators scale exponentially in the number of trainable parameters compared to the linear scaling of the quantum generator capacity, and that the classical generator capacity also scales exponentially. It means that for comparable performances, not only the classical generator needs to be exponentially more complex than the quantum generator, but also that the discriminator structure needs also to be exponentially more complex to achieve efficient training with high quality output. This hints towards a type of exponential advantage of the quantum generator for a latent style-based hybrid QGAN.

The paper is organized as follows. We introduce in Section~\ref{sec:methods} the computing framework, starting with GANs in Section~\ref{sec:methods:gans} and introducing WGAN-GP, then presenting the latent hybrid classical-quantum setup in Section~\ref{sec:methods:qgans} and our set of metrics in Section~\ref{sec:methods:metrics}. We present our results in Section~\ref{sec:results}, first introducing the details of chosen architecture for the experiments in Section~\ref{sec:results:structure} and the experimental setup in Section~\ref{sec:results:exp_setup} as well as the training procedure in Section~\ref{sec:results:training_procedure}, and then present and discuss the main results in Section~\ref{sec:results:exponential}, showing the exponential scaling of the classical discriminator and generator compared to the quantum generator. We also compare results obtained from the classical generator, the noiseless simulated quantum generator, and the sampled results from the IBM {\tt ibm_pittsburgh} quantum computer from a pre-trained model, in Section~\ref{sec:results:compare}. We conclude in Section~\ref{sec:conclusion}. 

\section{Methods}
\label{sec:methods}

\subsection{Generative Adversarial Networks}
\label{sec:methods:gans}

Generative Adversarial Networks (GANs) are formulated as a two-player minmax game between a generator $G$ and a discriminator $D$~\cite{goodfellow2014generativeadversarialnetworks}. The generator maps samples from a simple prior distribution $p(\boldsymbol{\xi})$ (e.g.\ Gaussian noise) to the data space, while the discriminator aims to distinguish real samples from generated ones.

Formally, GAN training is defined by the objective
\begin{equation}
\label{eq:gan}
\min_G \max_D \;
\mathbb{E}_{\mathbf{x} \sim p_{\mathrm{data}}}
\big[ \log D(\mathbf{x}) \big]
+
\mathbb{E}_{\boldsymbol{\xi} \sim p(\boldsymbol{\xi})}
\big[ \log \big(1 - D(G(\boldsymbol{\xi})) \big) \big],
\end{equation}
where $p_{\mathrm{data}}$ denotes the true (input) data distribution and $p(\boldsymbol{\xi})$ the noise distribution.

When (ideally) the Nash equilibrium is reached, the generator reproduces the data distribution and the discriminator cannot distinguish real from generated samples better than random guessing. While this formulation provides a principled framework for generative modeling with strong performance in high-quality image generation~\cite{10.1016/j.cag.2023.05.010}, medical image synthesis~\cite{karmakar2025role}, or molecular generation~\cite{kotkondawar2025generative}, its practical training often suffers from instabilities and sensitivity to model capacity and optimization dynamics, mode collapse, and a delicate balance between generator and discriminator capacity~\cite{heusel2017gans}. This motivates looking at the hybrid quantum-classical structure where the discriminator stays classical but the quantum generator is a parameterized quantum circuit (PQC).

Original implementations of GANs used the training procedure presented in Eq.~\ref{eq:gan}, leading to training issues linked to the non-continuous relation with the generator parameters. Using instead the Earth-Mover (or Wasserstein) distance overcomes this problem, and this was also refined into Wasserstein GAN with gradient penalty (WGAN-GP)~\cite{gulrajani2017improvedtrainingwassersteingans} where the penalty, and using gradient clipping as well. All these refinements mitigate the effect of mode collapse. The loss function of the discriminator becomes
\begin{equation}
    \mathcal{L} = 
\mathbb{E}_{\boldsymbol{\xi} \sim p(\boldsymbol{\xi})}
\big[D(G(\boldsymbol{\xi})) \big] -
\mathbb{E}_{\mathbf{x} \sim p_{\mathrm{data}}}
\big[D(\mathbf{x}) \big] + 
\lambda\, \mathbb{E}_{\boldsymbol{\hat{x}} \sim p(\boldsymbol{\hat{x}})}
\Big[\left(||\nabla_{\boldsymbol{\hat{x}}} D(\boldsymbol{\hat{x}})||_2 -1\right)^2 \Big],
\end{equation}
where $\boldsymbol{\hat{x}}$ are random sample drawn from $p(\boldsymbol{\hat{x}})$, sampling uniformly along straight lines between
pairs of points sampled from the input data distribution and the distribution of the generated output from the generator. Following \cite{gulrajani2017improvedtrainingwassersteingans}, a good value for the penalty term is $\lambda=10$. We will use the WGAN-GP formulation both for the classical and quantum GANs and from now on also uses the wording \textit{critic} to refer to the discriminator, following a standard formulation when employing WGAN-GP.

\subsection{Hybrid Quantum--Classical Latent GAN Architectures}
\label{sec:methods:qgans}

Despite their conceptual appeal, applying QGANs directly to high-dimensional data such as images remains challenging due to current hardware and simulation constraints. A common strategy to address this limitation is to introduce a classical dimensionality-reduction step that maps the data into a lower-dimensional latent space. Principal component analysis or autoencoders are typical choices for this purpose. Note that this latent-space strategy is also very valuable for classical GANs architecture, such that our classical benchmark will also use a latent approach.

Autoencoders (AE) encode the input data in a latent representation in a vector $\mathbf{z} \in \mathbb{R}^{d_z}$, where $d_z$ is the latent dimension, effectively learning the essential features of the dataset. The autoencoder $AE = Dec(Enc(\cdot))$ consists of two different networks, the encoder $Enc(\cdot)$ and the decoder $Dec(\cdot)$. The encoder learns how to embed the data into the latent space, therefore reducing the original data which is usually very high in dimension (like images) into a lower-dimensional latent vector $z$. The decoder network learns how to reconstruct the original data as close as possible based on the latent vector $z$, hence performing the reverse operation of the encoder.

In the hybrid latent QGAN architectures, the classical AE is first trained to compress the high-dimensional input data into the latent space. The generator and critic operate entirely in this latent space, and generated latent vectors are decoded back into the original data domain using the pre-trained decoder. This approach has been shown to significantly reduce the complexity of the generative task while preserving the semantic structure of the data~\cite{chang2024latentstylebasedquantumgan}.

\begin{figure}[t!]
    \centering
    \includegraphics[width=1\linewidth]{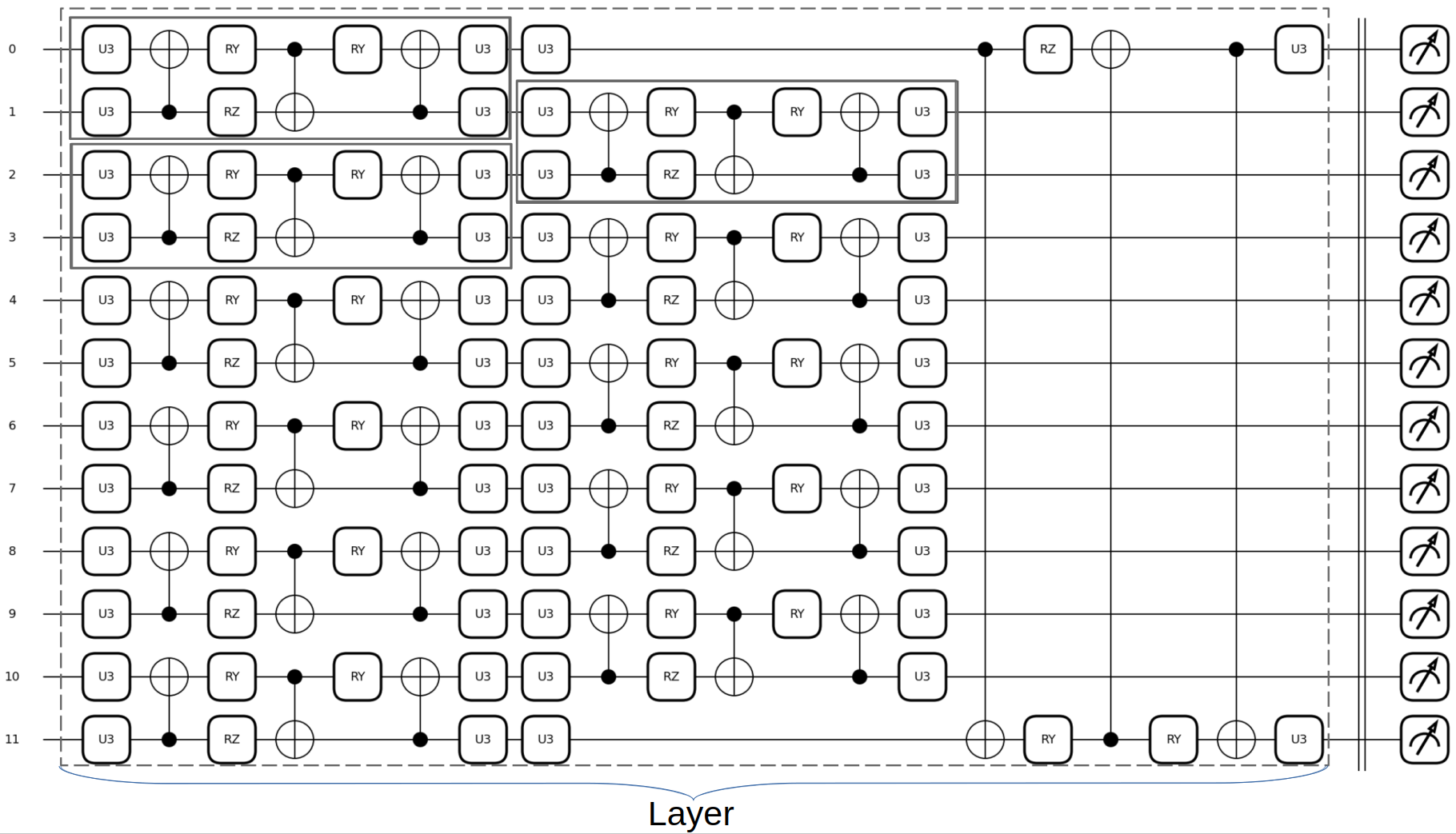}
    \caption{Variational quantum circuit with two-qubit-$SU(4)$ gates (gray bounded boxes) for one layer. Our study will use $L=2, 4, 6, 8$ layers.}
    \label{fig:circ3_ln}
\end{figure}

The quantum generator is implemented as a PQC using the style-based architecture~\cite{BravoPrieto2022}: All gates will contain input random noise. The ansatz we use is an $L$-repeating two-qubit-$SU(4)$-gate building block~\cite{chang2024latentstylebasedquantumgan, maccormack2022branching}(represented in gray boxes) entangling two neighboring qubits. One repetition (dashed box) of this quantum circuit is shown in Figure~\ref{fig:circ3_ln}. One $SU(4)$ box consists of $U3$ gates, which are essentially rotations around three axes so that any $U(3)$ gate is parameterized by three angles. One rotation angle in the domain $[0;2\pi]$ is defined as
\begin{equation}
    \theta_{q,\ell,k} = 2\pi \tanh\!\left( \xi_q \mathbf{W}_{q\ell k} + \mathbf{b}_{q\ell k}\right),
\end{equation}
where $\xi_q$ is the noise vector of the same dimension as the number of qubits $N_q$, $\mathbf{W}_{qlk}$ is the trainable parameters tensor for a given qubit $q$, a given layer $l$, and a given angle label $k$ representing the $SU(4)$ box. The tensor $\mathbf{b}$ contains the biases. We use a non-linear function $\tanh()$ to prohibit over-rotation during the training which will help to substantially mitigate quantum mode collapse.

One $SU(4)$ box contains four $U(3)$ gates (parameterized by three angles), two $R_Y$ rotations (parameterized by one angle), and one $R_Z$ rotation (parameterized by one angle), as well as 3 CNOT gates. Therefore, each $SU(4)$ box is parameterized by a total of $K=15$ angles, corresponding to a total of $N_{SU4}=30$ trainable parameters. As shown in Figure~\ref{fig:circ3_ln}, there are two consecutive repetitions of an $SU(4)$ box to entangle the blocks of two-qubit pairs together, so that in total a circuit of $L=2$ layers and $Q=12$ qubits contains $N_\Theta = N_{SU4}\times Q \times L = 30 \times 12 \times 2 = 720$ trainable parameters. 

\subsection{Benchmarks and Metrics}
\label{sec:methods:metrics}

As quantitative measurements for the generators image generation quality and diversity, we use the Fréchet Inception Distance (FID)~\cite{heusel2017gans}.  The FID measures the Fréchet distance between two multivariate Gaussian distributions, $\mathcal{N}(\mu_1, C_1)$ and $\mathcal{N}(\mu_2, C_2)$ and is defined as
\begin{equation}
\label{eq:fid}
\mathrm{FID} =
\left\lVert \boldsymbol{\mu}_1 - \boldsymbol{\mu}_2 \right\rVert_2^2 + \mathrm{Tr}\left(
\mathbf{C}_1 + \mathbf{C}_2 -
2\sqrt{\left(\mathbf{C}_1 \mathbf{C}_2\right)}
\right),
\end{equation}
where $\mu_n$ and $C_n$ denote the mean vector and covariance matrix of the respective distributions.

In practice, the two distributions correspond to the activations of a pretrained Inception-V3 network, extracted from the pool\_3 layer, for real samples and generated samples, respectively. These activations are treated as samples from multivariate Gaussian distributions, and their empirical means and covariances are computed over the evaluation set.

The dimensionality of the Inception pool\_3 feature space is 2048. Consequently, the number of samples used to estimate the Gaussian statistics must exceed this dimensionality to ensure that the covariance matrices are full rank. If this condition is violated, the matrix square root in Eq.~\eqref{eq:fid} may become ill-defined, leading to numerical instabilities such as complex values or NaNs.

To obtain reliable and numerically stable FID estimates, we follow standard practice and compute FID using at least 10,000 samples for both real and generated distributions. Using smaller sample sizes tends to systematically underestimate the true FID of a generator and increases variance in the metric.

To measure the reconstruction quality of the AE in a comparable way to the GAN, we define the reconstruction FID (\textbf{rFID}) for the AE, where distributions of an original evaluation dataset with $10,000$ samples is compared to the distribution of its own AE reconstructed images. We consider the rFID of the AE as a baseline, on which the GAN FID can be compared to since we have seen experimentally, that the quality of the AE has a big influence on the FID value of the GAN and it shows if there is still room to improve for the GAN if it does not reach the rFID. We use the {\tt Torch} implementation of the FID with N-Dims=2048~\cite{Seitzer2020FID}. 

We additionally evaluate the similarity between real and generated data distributions using the Jensen–Shannon Divergence (JSD)~\cite{jsd1991}. JSD is a symmetric and bounded measure of divergence between two probability distributions and is defined as
\begin{equation}
\label{eq:jsd}
\mathrm{JSD}(P\lVert Q) = \frac{1}{2}\mathrm{KL}(P\lVert M) + \frac{1}{2}\mathrm{KL}(Q\lVert M) \quad
\text{with} \quad M = \frac{1}{2}(P + Q), 
\end{equation}
where $KL(\lVert)$ denotes the Kullback–Leibler divergence. In all cases, P and Q denote empirical distributions estimated from finite sets of real and generated samples, either in pixel space or in feature space, depending on the variant of JSD reported \cite{weng2019gan}.

In practice, we estimate JSD between empirical distributions obtained from real and generated samples. We report two variants: The image-space JSD ($JSD_{raw}$) computed directly on the pixel distributions of real and generated images and the feature-space JSD ($JSD_{feat}$) computed on the latent vector $z$ from the AE.

While image-space JSD is sensitive to low-level differences in pixel statistics, it is highly affected by noise, resolution, and alignment artifacts, and thus often overestimates perceptual differences. Feature-space JSD mitigates these effects by comparing distributions in a semantically meaningful embedding space, but inherits the inductive biases of the pretrained network and, like FID, reflects similarity only up to the chosen representation.

Unlike FID, which assumes Gaussianity and matches only first- and second-order statistics, JSD compares full empirical distributions and therefore captures higher-order discrepancies. However, its estimation requires discretizations or density approximation, making it more sensitive to binning choices and sample size. Consequently, we treat JSD as a complementary diagnostic rather than a primary quality metric.

\pagebreak
\section{Results}
\label{sec:results}

\subsection{Structure of our latent style-based QGAN and of the classical GAN}
\label{sec:results:structure}

\begin{figure}[t!]
        \centering
        \includegraphics[width=\linewidth]{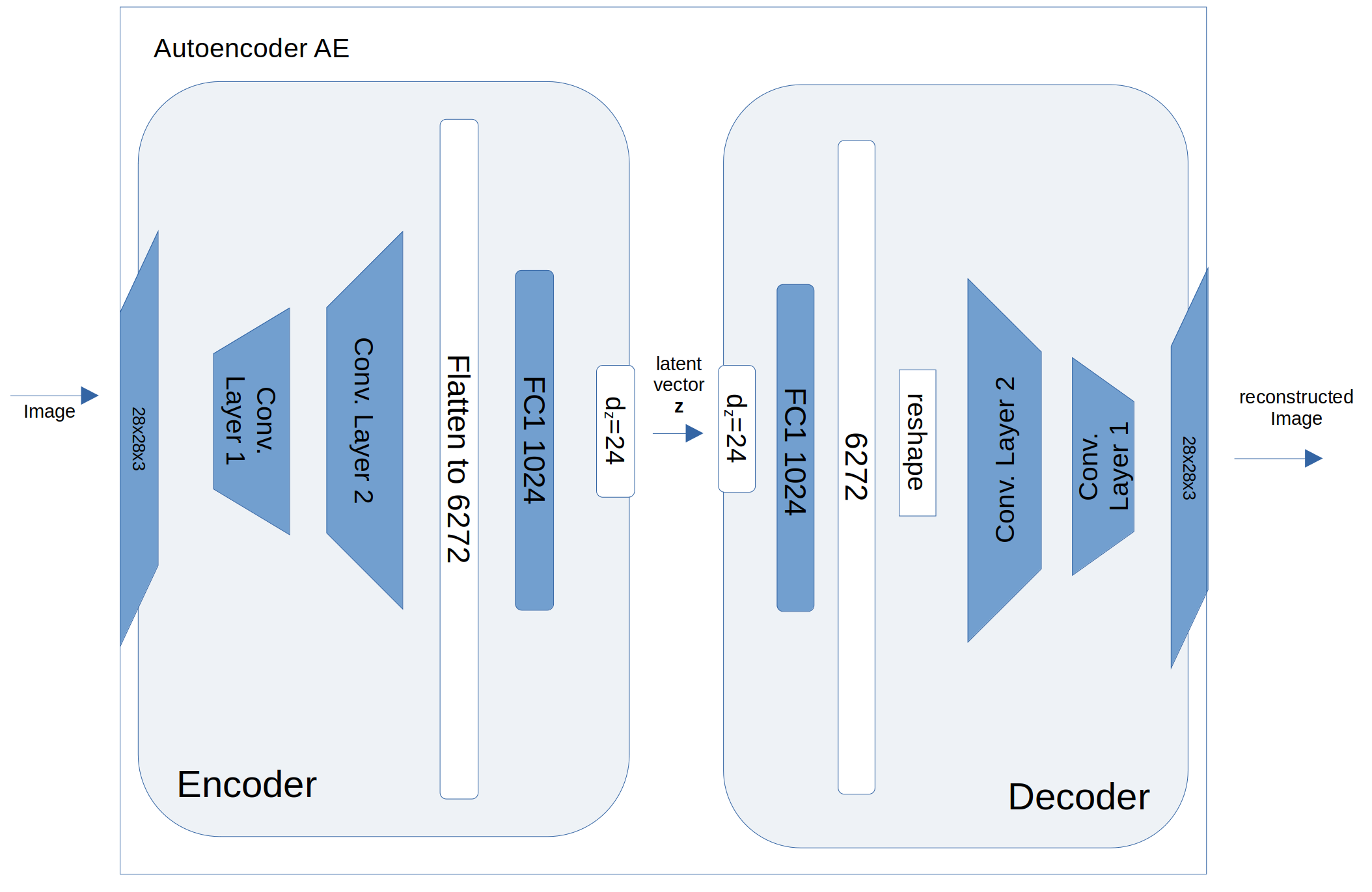}
        \caption{High-level autoencoder schematics, showing how the convolutional encoder compresses a 28x28x3 image into a latent vector of dimension 24, and then back up to a reconstructed image.}
        \label{fig:autoencoder}
\end{figure}

We use a classical AE based on a convolutional neural network, a classical critic also fully classical operating on latent vectors, and we compare classical and quantum versions of the generator. This setup allows for a controlled comparison between classical and quantum generators under identical training dynamics and evaluation metrics. The latent dimension and the corresponding trained AE are fixed across our experiments with the (Q)GANS, so that the comparison of generator and critic capacities ensuring stable adversarial training is not biased by potential different choices for the AE. We use the value $d_z=24$ for the latent dimension.

The AE is represented in Figure~\ref{fig:autoencoder}.It consists of two convolutional layers (with channel progressions of  Conv Layer 1 and 2 of $[64,128] $) to learn the features of the images, a single fully connected layer (FC width of 1024), which reduces the flattened output of the last convolutional layer into the dimension of the latent vector $z$ width size of $d_z=24$. The decoder is the opposite configuration: the input dimension is $d_z$, a fully connected layer connects into the first deconvolutional layer, which leads into the second deconvolutional layer, which in turn outputs the final image. Between the single layers of the encoder and decoder there are batch-normalization layers and dropout layers. Dropout layers randomly forget single values of the neural network, ensuring a more robust training and avoiding the tendency of large neural networks of simply replicating the input data, effectively learning it ``by heart''. 

\begin{figure}[t!]
    \centering
    \includegraphics[width=\linewidth]{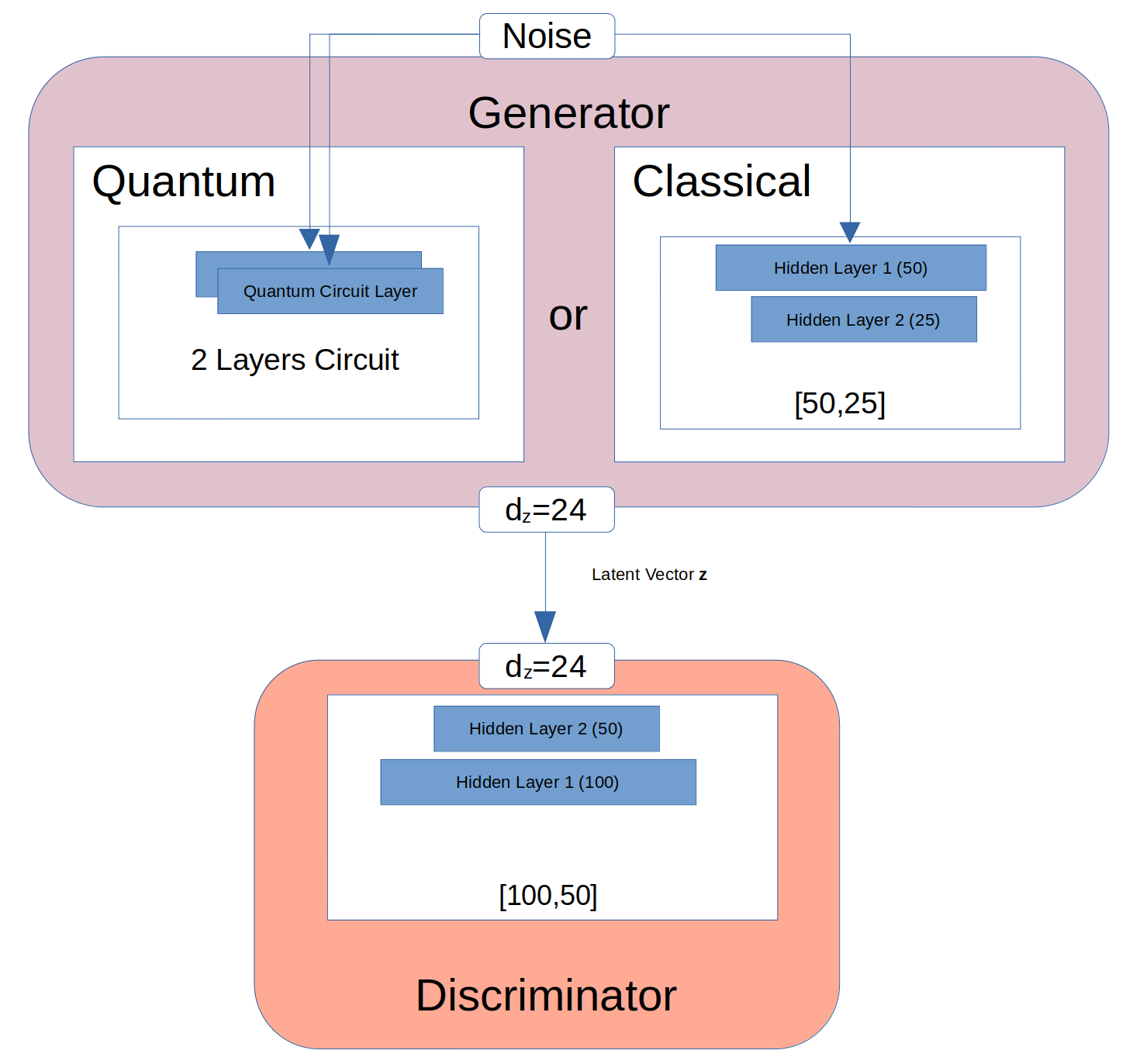}
    \caption{High-level GAN overview showing that for the generator we do either use a quantum generator with $L$ layers, or a classical deep neural network based generator with two layers, consisting with a width of $[N_1, N_2]$.}
    \label{fig:gan}
\end{figure}

The classical generator and classical critic are two-layer (dense) neural networks $L_{NN}=2$, where the width of each layer $N_{1,2}$ is denoted as $[N1, N2]$ and usually the layers are structured, such that $N_2 = \frac{N_1}{2}$. The total number of trainable parameters of such a deep neural network is denoted as $P$. Figure \ref{fig:gan} shows our integrated framework, where the quantum generator can directly be replaced with a classical generator counterpart. The total number of trainable parameters $P_{CG}$ for a classical generator with $L_{1,2}=[1400,700]$ is $P_{CG}=\num{1016401}$. Similarly, the total number of trainable parameters for the critic is $P_{D}= \num{1160274}$ for a critic of layers of sizes $N_{1,2}=[1500, 750]$. 

The style-based quantum generator consist of a PQC, as specified in Section~\ref{sec:methods:qgans} and depicted in Figure~\ref{fig:circ3_ln}. Using dual measurements, we only need $Q=12$ qubits for a latent dimension $d_z=24$, by measuring each qubit in the $X$ and in the $Z$ bases, respectively. We perform our experiments with a number of layers $L\in{2,4,6,8}$. The number of trainable parameters of the quantum generator $P_{QG}$ is then a linear function of the number of layers, with 720 parameters per layer $L$. We point out that while we use a style-based approach for the quantum generator (the input noise is injected in every gates in every layer), our classical generator counterpart does not have a style-based architecture and the noise is used only as an input in the first layer of the neural network.

\subsection{Experimental Setup}
\label{sec:results:exp_setup}

The hypothesis is that for each quantum generator configuration (in terms of circuit design and number of layers, which together result in a total number of trainable parameters), there is a matching classical critic configuration, such that the training of both the generator and the critic is in a balance and allows for a long-term stable training without over-powering behavior. This feature is essential to avoid a training break, where the GAN is not trainable anymore.

In our WGAN-GP two hyperparameters are fixed: the number of time the critic is updated for each cycle of generator training, $n_{\mathrm{critic}}=5$ or 10, and the gradient penalty term $\lambda_{gp}=1$. Only the critic capacity is changed, namely the width of the two layers in the neural network, as to reach our goal of stable training and stable low FID for a given capacity of the quantum generator. We use the Adam optimizer~\cite{kingma2015} for the training of both the AE and the (Q)GANs.

There are three different runs $R_1, R_2, R_3$ for three different seed numbers (42, 693094, 13671417). These seeds control the dataset sub-selection, the initialization of the neural network parameters, and as input for the random number generators used to sample the noise of the generator. To avoid any form of data leakage through the AE, the AE was trained on the exact same dataset sub-selection which is then used for the GAN training run.

The following table \ref{tab:exp_experiment_q} displays the tested configurations to find the optimal configuration for a given number of layers. 

\begin{table}[ht!]
    \centering
    \begin{tabular}{|c|c|c|c|c|c|}
    \hline
      Gen. Layers $L$    & $P_{QG}$ & min $D(\cdot)$ $L_{1,2}$ & min $P_{D}$ & max $D(\cdot)$ $L_{1,2}$ & max $P_{D}$\\
      \hline
       2  &  720 & [75,36] & \num{4638} & [350,175] & \num{70351} \\
       4  & 1440 & [125,62] &\num{11000} & [800,400] & \num{340801} \\
       6  & 2160 & [350,175] & \num{70351} & [900,450] & \num{428401} \\
       8  & 2880 & [1000,500] & \num{526001} & [1500,750] & \num{1321601}\\
         \hline
    \end{tabular}
    \caption{Experiment description for finding the optimal critic config for a given quantum generator configuration}
    \label{tab:exp_experiment_q}
\end{table}

The fixed hyperparameters for those experiments are given in Table~\ref{tab:hyperparameters}.

\begin{table}[ht!]
    \centering
    \begin{tabular}{cccc}
        Hyperparameter & Autoencoder $AE(\cdot)$ & Generator $G(\cdot)$ & critic $D(\cdot)$ \\
        \hline
        Latent dimension $d_z$ & 24 & 24 & 24\\
        Conv. Channel progression & [64,128] & - & - \\
        Linear Layers & 1024 & $L\in{2,4,6,8}$ & see tables \ref{tab:exp_experiment_q}, \ref{tab:exp_experiment_c}\\
        Dropout Rates & 0.0 & 0.0 & 0.0 \\
        Epochs & 100 & \num{10000} & \num{10000} \\
        Learning Rate (LR) & 0.001 & 0.0005 & 0.0008 \\
        $\beta_1$ & 0.9 & 0.5 & 0.5\\
        $\beta_2$ & 0.999 & 0.999 & 0.999\\
        Weight decay & 0 & 0 & 0 \\
        Dataset size & 2500 & 2500 & 2500\\
        Batch size & 12 & 256 & 256\\
        Initialization & Lecun Normal & Lecun Normal & Lecun Normal \\
    \end{tabular}
    \caption{Fixed configuration parameters for the AE, the GAN generator, and the GAN critic.}
    \label{tab:hyperparameters}
\end{table}

We have repeated the same setup for the experiments using a classical GAN. For a specific run, we pick the best performing critic for a given quantum generator (as a result of the experiments described above) and we explore a big range of classical generator capacities to find the best matching classical counterpart. The hyperparameters are the same as in Table~\ref{tab:hyperparameters}, except for the learning rates for the critic and the classical generator which have been changed to $LR_{D} = 0.0005$ and $LR_{CG}=0.0001$, respectively.

\begin{table}[ht!]
    \centering
    \begin{tabular}{|c|c|c|c|c|c|}
    \hline
      Gen. Layers $L$   & min $G_c(\cdot)$ $L_{1,2}$& min $P_{CG}$ & max $G_c(\cdot)$ $L_{1,2}$ & max $P_{CG}$. \\
      \hline
       2  &  [50,25] & \num{2449} & [400,200] & \num{89424} \\
       4  & [100,50] & \num{7374} & [700,350] & \num{261474} \\
       6  &  [300, 150] & \num{52074} & [1200, 600] & \num{748224} \\
       8  & [600, 300] & \num{194124} & [2000, 1000] & \num{2047024}\\
         \hline
    \end{tabular}
    \caption{Experiment description for finding the optimal classical generator configuration for a fixed critic, the latter being the best match for a given quantum generator.}
    \label{tab:exp_experiment_c}
\end{table}

We have performed noiseless quantum simulations in order to get our capacity scaling. The software to run these experiments has been written using {\tt Jax} / {\tt Flax}, allowing for an important speed-up in training time compared to a {\tt Torch} implementation, thanks to the optimized usage of graphical processing units (GPU). The quantum circuits have been coded with the Python package {\tt Pennylane} using its {\tt Jax} interface, allowing to also run the quantum circuit simulation on a GPU using the {\tt default.qubit} device and a backpropagation method to calculate the gradients. This setup has allowed for a massive reduction in training time on the quantum simulator, from days to hours: Training 10k epochs takes approximately 6 hours on a cloud virtual machine with an RTX6000 ADA GPU for a two-layer quantum simulation, and roughly 10 hours for an eight-layer quantum simulation, both with 12 qubits for $d_z=24$.

We have also tested the sampling of a trained QGAN on an IBM {\tt ibm_pittsburgh} quantum computer with a Heron R3 chip. A pretrained eight-layer quantum generator has been loaded on the quantum hardware and we have sampled new latent vectors to obtain 10,000 new images. We have fixed the number of shots to $n_{shots}=4000$ and used readout error mitigation (resilience level of 1 using IBM qiskit runtime). We have used 100 primitive unified blocks (PUBs) per submitted job, allowing for a sampling of 50 latent vectors per job as we need two measurements for each latent vector: one in the $X$ basis and a second in the $Z$ basis. In total, it amounts to 200 jobs, each running approximately 2 minutes on the quantum computer.

\subsection{Training Procedure}
\label{sec:results:training_procedure}

In our study, the AE is trained separately from the GAN, using the mean-squared error taken from the original image and the reconstructed image. To monitor the overfitting of the AE training, the reconstruction FID (rFID, see Section~\ref{sec:methods:metrics}) score between the original images from the validation dataset and their AE reconstructed images is calculated. The FID score is minimized when the distribution of the diversity and details of the features of images are matching.

The critic $D(\cdot)$ is trained to maximize the Wasserstein distance between real latent vectors $\mathbf{z}_{\mathrm{real}}$ obtained from the AE encoder and fake latent vectors $\mathbf{z}_{\mathrm{fake}} = G(\boldsymbol{\xi})$ generated from noise $\boldsymbol{\xi}$. The critic loss is defined as
\begin{equation}
\mathcal{L}_{\mathrm{D}}
=
\mathbb{E}_{\mathbf{z}_{\mathrm{fake}}}
\big[ D(\mathbf{z}_{\mathrm{fake}}) \big]
-
\mathbb{E}_{\mathbf{z}_{\mathrm{real}}}
\big[ D(\mathbf{z}_{\mathrm{real}}) \big]
+
\lambda_{\mathrm{gp}} \, \mathcal{L}_{\mathrm{GP}},
\end{equation}
where $\lambda_{\mathrm{gp}}$ is the gradient penalty coefficient.

The gradient penalty enforces the Lipschitz constraint and is defined as
\begin{equation}
\mathcal{L}_{\mathrm{GP}} =
\mathbb{E}_{\hat{\mathbf{z}}}
\Big[
\big(
\lVert \nabla_{\hat{\mathbf{z}}} D(\hat{\mathbf{z}}) \rVert_2 - 1
\big)^2
\Big],
\end{equation}
with $\hat{\mathbf{z}}$ sampled uniformly along straight lines between real and generated latent vectors.

The generator is trained to minimize the negative critic score of generated samples,
\begin{equation}
\mathcal{L}_{\mathrm{G}} =
- \mathbb{E}_{\boldsymbol{\xi} \sim p(\boldsymbol{\xi})}
\big[
D(G(\boldsymbol{\xi}))
\big].
\end{equation}
For each training step of one epoch, the critic is updated $n_{\mathrm{critic}}$ times, followed by $n_{\mathrm{gen}}$ generator updates. The usual GAN setup follows $n_{\mathrm{critic}}=5$ and $n_{\mathrm{gen}}=1$. As we want to test some variations we have implemented the possibility of $n_{\mathrm{gen}}>1$ even if this is rarely used, see Appendix~\ref{appendix:pseudocode}.

In addition to the WGAN-GP objective, we have applied gradient norm clipping as an optimizer-side stabilization throughout all experiments. Let $\mathbf{g}$ denote the gradient of the current objective with respect to the model parameters. Before applying the optimizer update, gradients are rescaled as
\begin{equation}
\mathbf{g} \leftarrow \mathbf{g}\cdot \min\left(1,\frac{c}{\lVert \mathbf{g}\rVert_2}\right),
\qquad c = 5,
\end{equation}
which bounds the global gradient norm by $c$. This operation is applied to both generator and critic updates unless stated otherwise. We include this mechanism to prevent rare large gradient events from destabilizing training. It is thus a conservative safeguard on top of the gradient penalty term. We have not performed a dedicated ablation of gradient clipping: we report it for completeness and reproducibility.

\subsection{Main result on the exponential capacity scaling}
\label{sec:results:exponential}

\begin{figure}[t!]
        \centering
        \includegraphics[width=1\linewidth]{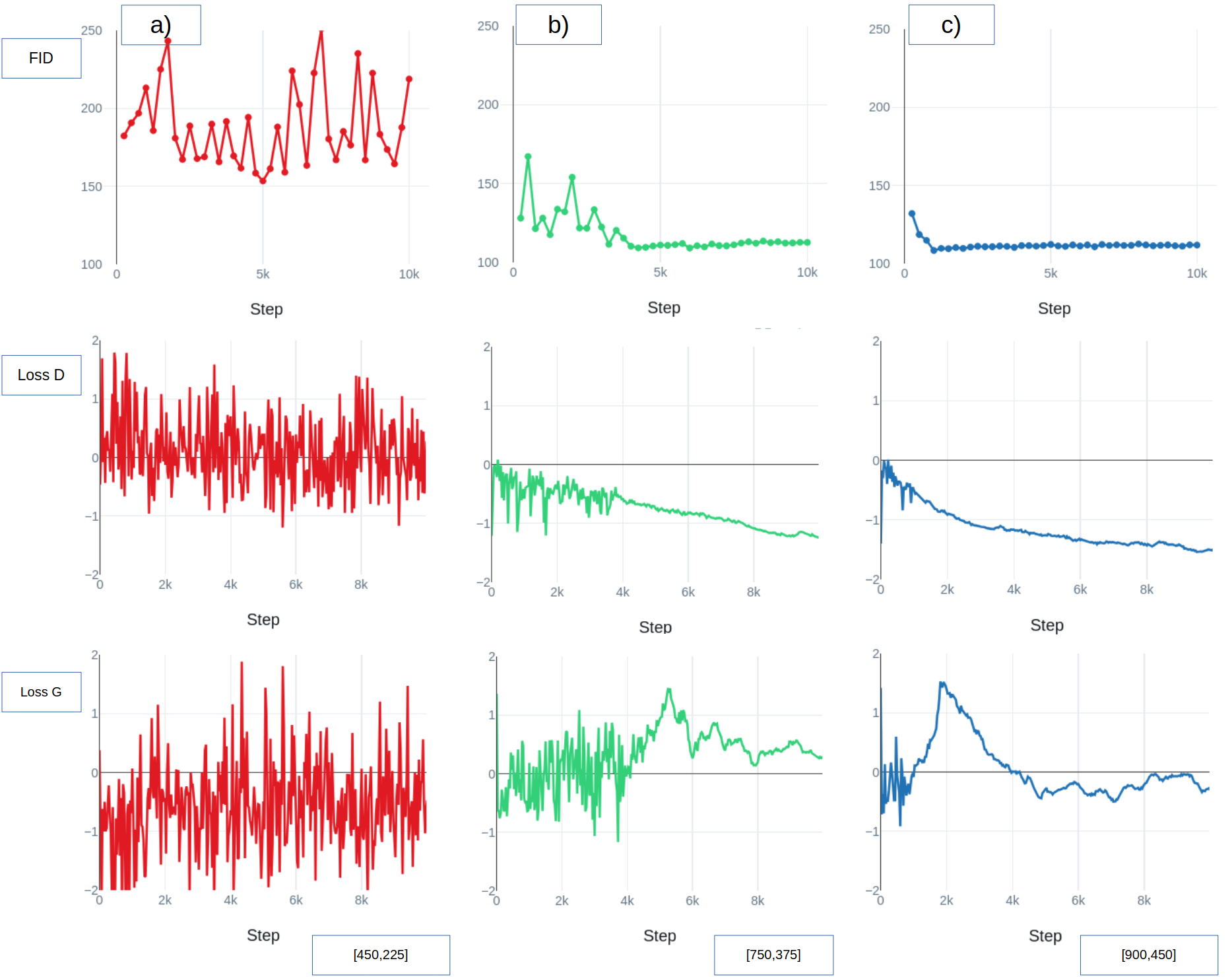}
        \caption{Columns a), b) and c) each contain figures for different critic capacities $L_{1,2}$ for a run with a quantum generator of $L=6$ layers. The first line displays validation FID scores (FID), the second line displays the critic training losses (Loss D), and the third line displays the generator training losses (Loss G).}
        \label{fig:disc_config_selection}
\end{figure}

We present the results of our experiments establishing the exponential capacity scaling of both the critic and the classical generator compared to the quantum generator.
We display in Figure~\ref{fig:disc_config_selection} the training losses and the FID evolution as a function of the number of training steps, for three critic capacities. For small critics, columns a), the FID evolution is not stabilized at all and the training is chaotic. For very large critics, on the contrary, the FID is very stable and reaches a plateau quite early on in the training, while the loss functions are stable and smooth as seen in column c). The transition is depicted in column b) where the training is still chaotic until a phase transition happens at around 4k steps. From capacities of the type Figure~\ref{fig:disc_config_selection} b) to the big capacities of the type Figure~\ref{fig:disc_config_selection} c), there is no much change in FID value: the best FID is thus obtained very close to this capacity transition.

This transition point is our target to define the optimal capacity of the discriminator: we avoid overfitting with too large a discriminator for a given quantum generator, while ensuring stable training and smooth stable low value for the FID score. We obtain for each run this optimal point, and draw the number of trainable parameters (capacity) for the optimal critic, $P_D$, as a function of the corresponding quantum generator capacity, $P_G$. We can repeat the same exploration for the classical generator as well.

\begin{figure}[t!]
    \centering
    \includegraphics[width=\linewidth]{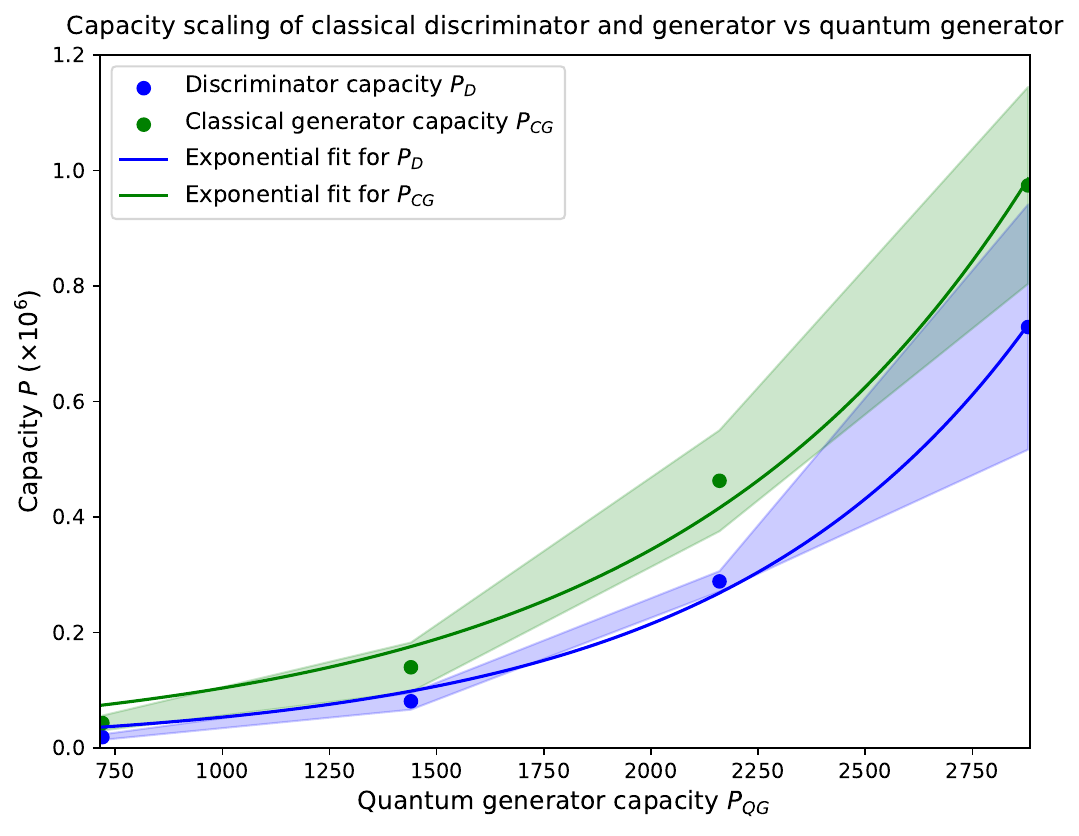}
    \caption{Capacity scaling of the best performing discriminator (critic) $P_D$ (blue) and of the the best performing classical generator $P_{CG}$ (green) as a function of the quantum generator trainable parameters $P_{QG}$. We also display the $1\sigma$ bands of the capacity of the classical critic and generators, as well as the corresponding exponential fits of the data points (solid lines).}
    \label{fig:exp_capacity}
\end{figure}

The result is displayed in Figure \ref{fig:exp_capacity} where the critic capacity scaling is displayed in blue together with an exponential fit (blue line), while the classical generator capacity scaling is displayed in green with the corresponding exponential fit (green line). In both case we obtain a clear exponential capacity scaling, meaning that the critic and the classical generator need to become exponentially large to match the capability of the style-based quantum generator. The blue and green bands represent the $1\sigma$ deviation calculated over our three different initial random seeds. The exponential fit function reads $P = a*e^{b*T_{QG}}$, where $P$ is the capacity of the critic or of the classical generator, $a,b$ are the fitting coefficients, and $P_{QG}$ is the capacity of the quantum generator (ranging from 720 for two layers to 2880 for eight layers). While it has been hinted several time in the literature that quantum GANs can perform well with a significantly lower number of trainable parameters compared to classical GANs, Figure~\ref{fig:exp_capacity} presents an experimental systematic evidence of an exponential power of quantum circuits.

\subsection{Comparative FID analysis of generated images from classical GAN, analytical style-based QGAN and from quantum hardware runs}
\label{sec:results:compare}

So far we have used a noiseless quantum simulation to train and sample our latent style-based QGAN, assuming infinite shots. We present in this section sampling results when running the noiseless pre-trained QGAN on an IBM quantum computer and compare to both the output from the AE, the classical GAN, and the noiseless quantum simulated style-based QGAN. We select our best runs from the experiments we have performed and the sampling on real hardware is performed on the IBM {\tt ibm_pittsburgh} quantum computer. As displayed in Table~\ref{tab:qsim_best} the average FID score for 10,000 generated SAT4 images using a noiseless quantum simulator is only 5 points above the baseline reconstruction FID from the AE, 109.6 vs 104.7. The same comparison is performed for the best classical runs and is displayed in Table~\ref{tab:classic_best}. The average classical FID score is now slightly better than the reconstruction FID, 104.0 vs 104.7. Comparing classical and quantum GANs, we get a 5~\% degradation in performance from the quantum GAN, which is quite reasonable. Note again that the average capacities of the classical neural network architectures are much larger than the quantum generator capacity.

\begin{table}[ht!]
    \centering
    \begin{tabular}{ccccccccc}
        Exp. run & $L$ & $P_{QG}$ & $D(\cdot) L_{1,2}$ & $P_D$ & FID & JSD raw & JSD feat & AE rFID \\
        \hline
        $R_1$      & 8 & 2880 &[1400,700] &\num{1016401} & 109.4 & 1.77 & 0.66 & 103.3 \\
        $R_2$  &  8 &2880 & [1400,700] &\num{1016401} & 110.6 & 1.19 & 0.20 & 107.5\\
        $R_3$ &  8 &2880 & [1100, 550] & \num{633601} &108.7 & 1.30 & 0.36 & 103.2  \\
        \hline
        Average & 8 & 2880 & - & \num{888801}  & 109.6 & 1.42 & 0.40 & 104.67 \\
    \end{tabular}
    \caption{List of the best sampling runs using a noiseless quantum simulator for an eight-layer quantum circuit. $L$ stands for the number of layers, $P_{QG}$ for the capacity of the quantum generator, $P_D$ for the capacity of the critic. We display the critic structure as well as the obtained FID, JSD, and AE reconstruction FID (rFID). On average the quantum FID is 5 points bigger than the rFID.}
    \label{tab:qsim_best}
\end{table}

\begin{table}[ht!]
    \centering
    \begin{tabular}{ccccccccc}
        Exp. run & $G(\cdot)L_{1,2}$ & $P_{CG}$ & $D(\cdot)L_{1,2}$ & $P_D$ & FID & JSD raw & JSD feat & AE rFID \\
        \hline
        $R_1$     & [600, 300] & \num{194124} & [750, 375] & \num{300751} & 104.3 & 0.37 & 0.57 & 103.3 \\
        $R_2$  &  [1500, 750] & \num{1160274} & [1400,700] & \num{1016401} & 103.6 & 0.18 & 0.30 & 107.5\\
        $R_3$ &  [400, 200] & \num{89424} & [350, 175] & \num{70351} &104.0 &  0.57 & 0.22 & 103.2  \\
        \hline
        Average & - & \num{481274} & - & \num{462501}  & 104.0 & 0.37 & 0.36 & 104.67 \\
    \end{tabular}
        \caption{Same as in Table~\ref{tab:qsim_best} but for the classical GAN and where $P_{CG}$ stands for the capacity of the classical generator. The average classical FID is slightly better than the reconstruction FID.}
    \label{tab:classic_best}
\end{table}

How does the (best) classical GAN perform when its capacity becomes comparable to that of the quantum generator? To answer this question we have selected the runs with the smallest capacity for the classical generator, $P_{CG}=2449$ trainable parameters, corresponding to a critic capacity of $P_D=25,000$ trainable parameters. This order of magnitude corresponds to a quantum generator with $P_G=2880$ trainable parameters as selected in Table~\ref{tab:qsim_best}. The results are displayed in Table~\ref{tab:classic_small} and demonstrate that the FID score of the classical GAN is now close to that of the style-based quantum generator, 107.8 vs 109.6. As $P_G=2880$ is the largest quantum generator capacity we have tested in our experiments, it is quite likely that there is still room for improvement for the quantum generator and that the best FID we can get is for larger quantum generator capacities. Following our results in the previous section, this also means that the best FID from the quantum generator will be on par, if not better, than the AE reconstruction FID and the classical FID, but using exponentially far less trainable parameters than the corresponding classical critic and generator.

\begin{table}[ht!]
    \centering
    \begin{tabular}{ccccccccc}
        Exp. run & $G(\cdot) L_{1,2}$  & $P_{CG}$ & $D(\cdot) L_{1,2}$ & $P_D$ & FID & JSD raw & JSD feat & AE rFID \\
        \hline
        $R_1$ & [50, 25] & 2449 & [200, 100] & \num{25201} & 108.6 & 0.94 & 0.10 & 103.3 \\
        $R_2$ & [50, 25] & 2449 & [200, 100] & \num{25201} & 107.0 & 0.41 & 0.09 & 107.5\\
        $R_3$ & [50, 25] & 2449 & [200, 100] & \num{25201} & 107.9 & 1.06 & 0.30 & 103.2  \\
        \hline
        Average & [50, 25] & 2449 & [200, 100] & \num{25201}  & 107.8 & 0.80 & 0.16 & 104.67 \\
    \end{tabular}
    \caption{Same as in Table~\ref{tab:classic_best} but for a fixed classical generator capacity of $P_{CG}=2449$ trainable parameters, matching the order of magnitude of the quantum generator capacity, $P_G=2880$.}
    \label{tab:classic_small}
\end{table}

We finish this section with the results on the IBM {\tt ibm_pittsburgh} quantum hardware. We compare in Table~\ref{tab:quantum_sampling} the metrics for the sampling of $10,000$ SAT4 images when using the noiseless quantum simulator and the {\tt ibm_pittsburgh} quantum hardware with readout error mitigation on $4000$ shots (qiskit runtime resilience level 1). The error mitigation technique provided by IBM relies on measurement twirling, applying Twirled Readout Error eXtinction (TREX)~\cite{vandenBerg2022}. The strongest eight-layer quantum generator was pre-trained with the noiseless quantum simulator, and a checkpoint was taken at the best measured FID evaluation run. This checkpoint serves as input trainable parameters for the sampling on both the quantum simulator and on the quantum hardware. The quantum noise induced by the real quantum computer degrades the image quality by 17~\%, with an FID reaching 128.2 compared to 109.5 on the noiseless quantum simulator.

\begin{table}[ht!]
    \centering
    \begin{tabular}{cccccccccc}
        Exp. run & $G(\cdot) L$  & $P_{CG}$  &$D(\cdot) L_{1,2}$  &  $P_D$  & AE rFID  & Device & FID & JSD raw & JSD feat \\
        \hline
        $R_1$  &  8 & 2880 &[1400,700] &\num{1016401} &103.3 & noiseless simulator & 109.5 & 0.31 & 0.67  \\
        $R_1$  &  8 & 2880 & [1400,700] &\num{1016401} & 103.3 & {\tt ibm_pittsburgh} &128.2 & 3.15 & 0.28 \\
        
    \end{tabular}
    \caption{Comparison metrics for the sampling of 10k images from a noiseless quantum simulator and from the IBM {\tt ibm_pittsburgh} quantum computer with readout error mitigation and $4000$ shots.}
    \label{tab:quantum_sampling}
\end{table}

\begin{figure}[t!]
    \centering
    \includegraphics[width=\linewidth]{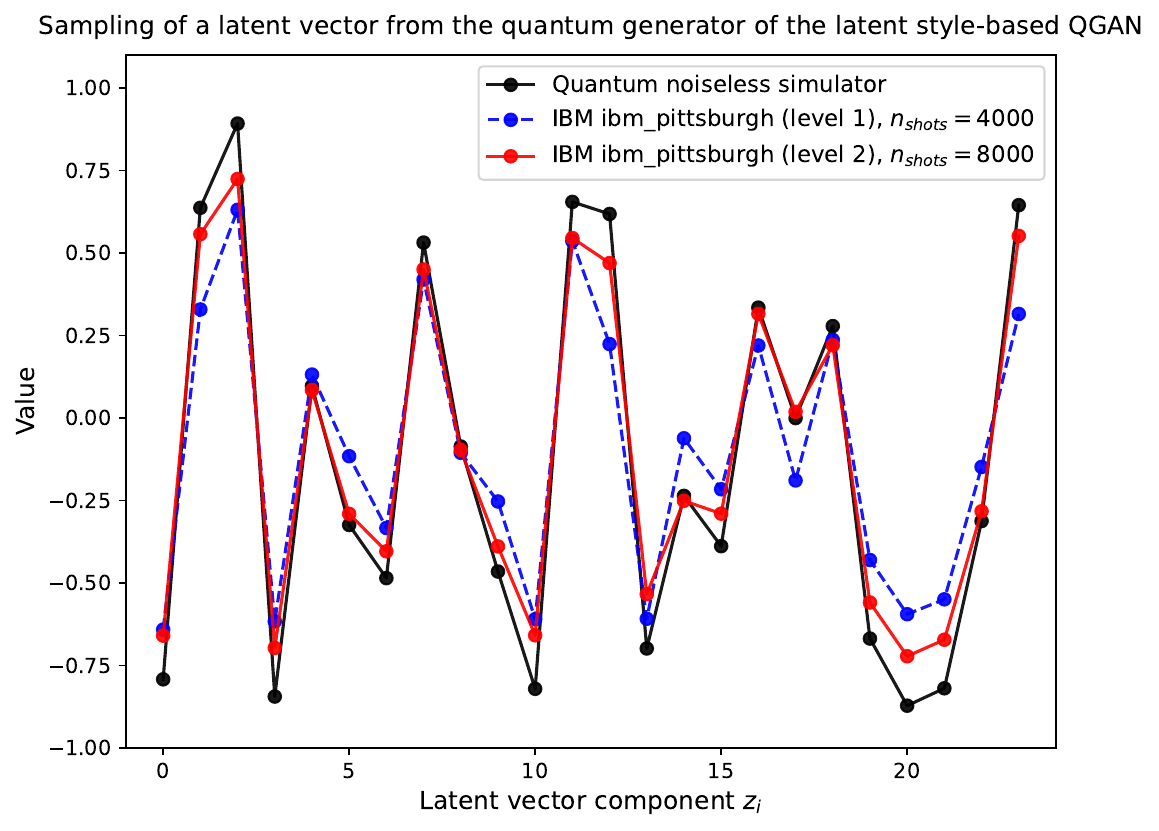}
    \caption{Comparison of three latent vectors generated by the quantum generator from the same noise input, sampled on: a) the noiseless quantum simulator (in solid black); b) the IBM {\tt ibm_pittsburgh} quantum hardware using readout error mitigation and $4000$~shots (level 1, in dashed blue); c) {\tt ibm_pittsburgh} using $8000$~shots, readout error mitigation, and ZNE (level 2, in solid red).}
    \label{fig:raw_latent_vector_comparison_quantum_sim}
\end{figure}

We have also performed a second run on the quantum hardware with a resilience level of 2, applying not only readout error mitigation but also gate twirling and the Zero Noise Extrapolation method (ZNE)~\cite{Temme2017}. At the level of the quantum generator output, in the latent space, we display in Figure~\ref{fig:raw_latent_vector_comparison_quantum_sim} three latent vectors sampled from the same noise input, by a) the quantum simulator; b) the quantum hardware with readout error mitigation (qiskit runtime resilience level of 1); c) the quantum hardware with readout error mitigation and ZNE on top (qiskit runtime resilience level of 2). The first quantum sampling with resilience level of 1 and 4000 shots has been used to sample the $10,000$ images for the FID evaluation in Table~\ref{tab:quantum_sampling}. The quantum runtime for one job generating 20 latent vectors is 54~seconds. Calculating the average mean squared error (MSE) over these 20 latent vectors compared to the noiseless quantum simulation, we obtain $\mathrm{MSE}_{z}=0.142$. We have used $8000$ shots for the second quantum sampling with resilience level of 2, to increase the accuracy of the results when comparing at the latent vector level. The runtime for one job generating 20 latent vectors is 4~minutes and 30~seconds. The overall MSE between these 20 latent vectors and their noiseless quantum-simulated sampled counterparts is $\mathrm{MSE}_{z}=0.078$, almost cutting the error by 50~\% and demonstrating the improvement obtained with both more shots (increasing the statistical accuracy) and advanced error mitigation techniques. 

Over the entire span of the latent dimension there are discrepancies between the three latent vectors which induce the FID difference we get in Table~\ref{tab:quantum_sampling}. Comparing the JSD raw scores indicates a degradation by a factor of 10, signaling a major reduction in the diversity of the sampled images, while the JSD value of the sampled feature improves slightly. The reason is likely to come from the additional source of noise induced by the quantum hardware, but as this noise is random and not related to the structure of the latent space (captured by the AE), it degrades the final output image quality and diversity. 

\begin{figure}[t!]
    \centering
    \includegraphics[width=1\linewidth]{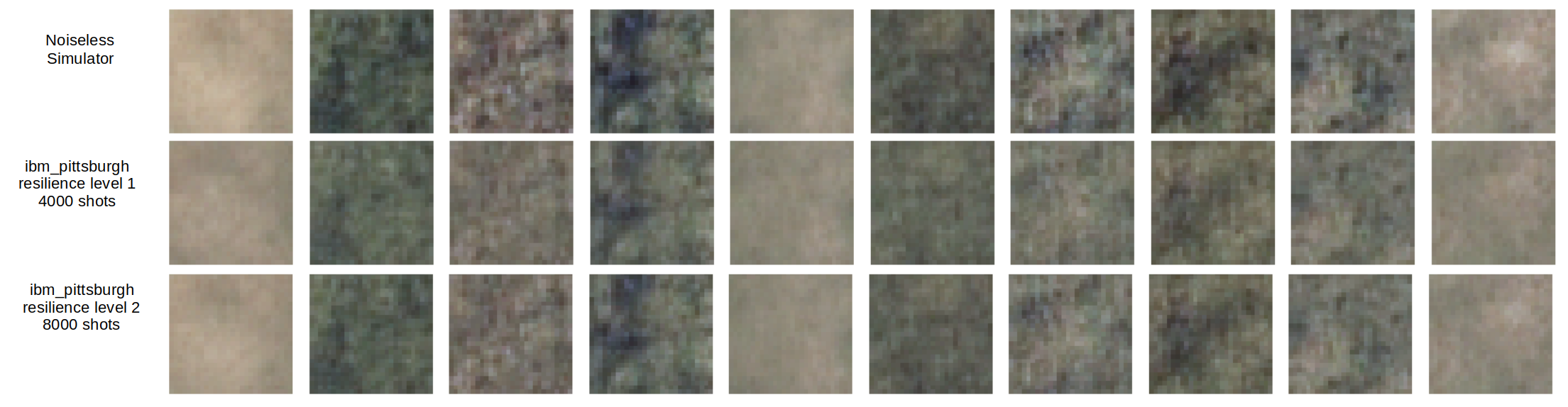}
    \caption{A one-to-one comparison of the sampling of 10 SAT4 images on the noiseless quantum simulator and the IBM {\tt ibm_pittsburgh} quantum computer, using the same noise vectors as input of the quantum generator in all cases. Sampling on the quantum computer was done with two different resilience modes (1 and 2) and numbers of shots (4000 and 8000), to reduce the impact of quantum noise.}
    \label{fig:image_comparison}
\end{figure}

We also provide in Figure~\ref{fig:image_comparison} a snapshot of 10 images sampled by our latent style-based QGAN, comparing on common grounds the output from the noiseless quantum simulator and from the {\tt ibm_pittsburgh} quantum hardware with the two different resilience levels and shots settings. The same noise vector is used as input of both runs, so that in the ideal case this should produce the same latent vector and hence the same image as the AE decoder is identical in both cases. As exemplified in Figure~\ref{fig:raw_latent_vector_comparison_quantum_sim} and Table~\ref{tab:quantum_sampling} there is degradation induced by quantum noise, however surprisingly enough this is barely visible for the human eye.

\section{Conclusion}
\label{sec:conclusion}

In this work, we have presented a systematic experimental study of capacity scaling for a latent Wasserstein GAN architecture using gradient penalty (WGAN-GP), comparing the classical latent WGAN-GP with a hybrid latent style-based quantum GAN (QGAN) where the critic stays classical while the generator is a quantum circuit. Using satellite image generation on the SAT4 dataset as a concrete and practically relevant benchmark, we have investigated how the number of trainable parameters, that are required for stable adversarial training and stable and high-quality (low) FID scores, scales across classical and quantum models.

Performing noiseless quantum simulations with 12 qubits for a latent dimension of 24, We have obtained an exponential separation in capacity scaling between classical and quantum generators within the hybrid latent style-based QGAN framework. Specifically, once the autoencoder (AE) used to embed the input satellite images in an abstract latent space of vectors of dimension 24 is trained and fixed, we have found that the optimal classical discriminator (in term of training stability and of stable low values of the FID score), corresponding to a given capacity of the quantum generator, scales exponentially with respect to this generator capacity. We have also obtained this exponential behavior for the classical generator of a classical latent WGAN-GP, where the critic is the same as the optimal critic for the latent style-based QGAN and the optimal classical generator also ensures stable training and stable lowest values of the FID score: the number of trainable parameters of the classical generator also scales exponentially with respect to the number of trainable parameters of the quantum generator. This result has been observed consistently across different random seeds initializing our experiments: different dataset sub-selections, independently trained AEs, different initialization of the trainable parameters. This provides strong evidence that the effect is robust and not an artifact of a particular training instance.

Beyond the scaling analysis, we have also observed that the absolute performance of the latent style-based QGAN is close to that of its classical counterparts. With a noiseless quantum simulation, the FID score of our latent style-based QGAN is only 5\% worse that our best classical WGAN-GP architecture, with only 2800 trainable parameters for the quantum generator compared to an average of 481,274 parameters for the classical latent WGAN-GP. When comparable capacities were chosen for the classical and quantum generators, the quantum FID scores was found to be less than 2~\% different from the best classical FID score. Given that this capacity for the quantum generator was the highest we tested, corresponding to eight layers, and given that the quantum generator capacity scales linearly with the number of layers, it hints to a regime where the latent style-based QGAN delivers FID scores on par if not better than the best classical latent WGAN-GP, while the capacity of the quantum generator only increases marginally and, according to our central result, the capacity of the classical generator and of the classical critic grow exponentially

We have also sampled $10,000$ images from the IBM {\tt ibm_pittsburgh} quantum hardware with two options, either using only readout error mitigation or using in addition Zero Noise Extrapolation (ZNE) to further increase the accuracy of the quantum hardware output and reduce the impact of quantum noise. The latter has been found to degrade image quality and diversity as expected, but the generated samples remain structurally meaningful and preserve latent-space features, highlighting the feasibility of executing such hybrid architectures on current quantum hardware.

A key methodological insight of this study is the critical role of the AE in latent GAN training. We have found that stable adversarial dynamics and meaningful capacity comparisons are only achievable once the AE reconstruction quality reaches a sufficiently low and stable baseline for the FID score. This observation applies equally to classical and quantum GANs and underscores the importance of treating the AE as an integral component of the overall generative pipeline rather than a mere prepossessing step.

Taken together, our results provide the first comprehensive experimental hints that hybrid quantum--classical generative models can exhibit a capacity advantage over fully classical architectures for image generation. While this does not yet constitute a practical end-to-end performance advantage in terms of final image quality, the observed exponential separation in required classical capacity strongly suggests that
quantum circuits can represent and explore generative distributions with significantly fewer trainable parameters without sacrifying the output quality.

Tere are several directions for future work. On the quantum side, exploring
alternative circuit ans\"atze, noise-aware training strategies, and improved encoding schemes may further narrow or close the remaining performance gap. On the classical side, a deeper theoretical understanding of why the discriminator capacity must scale exponentially in the
presence of a quantum generator may shed light on the fundamental representational differences between classical neural networks and parameterized quantum circuits. More broadly, this work establishes capacity scaling as a concrete and experimentally accessible lens through which potential quantum advantages in generative modeling can be studied.

\section*{Acknowledgments}

This work has been funded by ARTIDIS AG, which has provided financial support, classical GPU computational resources, and coverage for external consultancy and quantum hardware access. The authors thank in particular Marko Loparic for continuous support, strategic guidance, and alignment of this research with practical application use cases. QuantumBasel is acknowledged for IBM quantum hardware access as well as organizational and technical support throughout the project. We thank Janine Goldiger for exceptional organizational and management support. Parts of this work have been executed using IBM Quantum Services.  The views expressed in this work are those of the authors and do not necessarily reflect the views of IBM. The authors declare that there is no conflict of interest.

\appendix
\section{Notation and Abbreviations}
\label{app:notation}

This appendix summarizes all symbols, abbreviations, and indices used throughout the manuscript.
Unless stated otherwise, bold symbols denote vectors or tensors.

\subsection{Indices}

\begin{center}
\begin{tabular}{lll}
\hline
Symbol & Meaning & Range \\
\hline
$i$ & Sample index & $i = 1,\dots,N_{\text{samples}}$ \\
$q$ & Qubit index & $q = 1,\dots,Q$ \\
$\ell$ & Quantum circuit layer index & $\ell = 1,\dots,L$ \\
$k$ & $SU(4)$ internal angle index & $k = 1,\dots,15$ \\
$l$ & Classical NN layer index & $l = 1,\dots,L_{\mathrm{NN}}$ \\
\hline
\end{tabular}
\end{center}

\subsection{Model Architecture}

\begin{center}
\begin{tabular}{lll}
\hline
Symbol & Description & Example / Comment \\
\hline
$Q$ & Number of qubits & $Q=12$ \\
$L$ & Number of quantum circuit layers & $L \in \{2,4,6,8\}$ \\
$L_{\mathrm{NN}}$ & Number of layers in a classical NN & Generator / Critic \\
$N_l$ & Width of NN layer $l$ & e.g.\ $[600,300]$ \\
\hline
\end{tabular}
\end{center}

\subsection{Latent Space and Noise}

\begin{center}
\begin{tabular}{lll}
\hline
Symbol & Description & Notes \\
\hline
$\boldsymbol{\xi}$ & Noise vector (generator input) & $\boldsymbol{\xi} \sim p(\boldsymbol{\xi})$ \\
$\xi_q$ & Noise component for qubit $q$ & $d_\xi = Q$ \\
$\mathbf{z}$ & Latent vector (AE / GAN interface) & $\mathbf{z}\in\mathbb{R}^{d_z}$ \\
$d_z$ & Latent dimension & $d_z = 24$ \\
$d_\xi$ & Noise dimension & $d_\xi = Q$ \\
\hline
\end{tabular}
\end{center}

\subsection{Quantum Generator Parameters}

\begin{center}
\begin{tabular}{lll}
\hline
Symbol & Description & Notes \\
\hline
$\theta_{q,\ell,k}$ & Rotation angle & $SU(4)$ parameter \\
$\mathbf{W}_{q\ell k}$ & Trainable weight & Noise-to-angle mapping \\
$\mathbf{b}_{q\ell k}$ & Trainable bias & Angle offset \\
$N_{\theta}^{\mathrm{SU4}}$ & Angles per $SU(4)$ block & $15$ \\
$P_{\mathrm{SU4}}$ & Parameters per $SU(4)$ block & $30$ \\
\hline
\end{tabular}
\end{center}

The rotation angles are defined as
\begin{equation}
\theta_{q,\ell,k}
=
2\pi \tanh\!\left(
\xi_q \mathbf{W}_{q\ell k} + \mathbf{b}_{q\ell k}
\right),
\end{equation}
which ensures bounded rotations.

\subsection{Trainable Parameter Counts}

\begin{center}
\begin{tabular}{lll}
\hline
Symbol & Description & Usage \\
\hline
$P_{\mathrm{QG}}$ & Quantum generator parameters & $P_{\mathrm{QG}} = 30QL$ \\
$P_{\mathrm{CG}}$ & Classical generator parameters & NN-based \\
$P_{\mathrm{D}}$ & Discriminator parameters & NN-based \\
$P_{\mathrm{AE}}$ & Autoencoder parameters & Encoder + Decoder \\
$P_{\mathrm{tot}}$ & Total parameters & Optional \\
\hline
\end{tabular}
\end{center}

\subsection{Training Hyperparameters}

\begin{center}
\begin{tabular}{lll}
\hline
Symbol & Description & Typical value \\
\hline
$n_{\mathrm{critic}}$ & Critic updates per step & $5$ \\
$n_{\mathrm{gen}}$ & Generator updates per step & $1$ \\
$\lambda_{\mathrm{gp}}$ & Gradient penalty coefficient & WGAN-GP \\
$e$ & Epoch index & Training loop \\
$t$ & Training step index & Within epoch \\
\hline
\end{tabular}
\end{center}

\subsection{Losses and Metrics}

\begin{center}
\begin{tabular}{lll}
\hline
Symbol & Description & Notes \\
\hline
$\mathcal{L}_{\mathrm{D}}$ & Critic loss & Wasserstein + GP \\
$\mathcal{L}_{\mathrm{G}}$ & Generator loss & $-\mathbb{E}[D(G(\cdot))]$ \\
$\mathcal{L}_{\mathrm{W}}$ & Wasserstein loss & Raw separation \\
$\mathcal{L}_{\mathrm{GP}}$ & Gradient penalty & Lipschitz constraint \\
$\mathrm{FID}$ & Fréchet Inception Distance & Primary metric \\
$\mathrm{rFID}$ & Reconstruction FID & AE baseline \\
$\mathrm{JSD}$ & Jensen–Shannon divergence & Complementary metric \\
\hline
\end{tabular}
\end{center}

\subsection{Functions and Distributions}

\begin{center}
\begin{tabular}{lll}
\hline
Symbol & Description & Notes \\
\hline
$G(\cdot)$ & Generator function & Quantum or classical \\
$D(\cdot)$ & Critic & Scalar output \\
$Enc(\cdot)$ & AE encoder & Image $\rightarrow$ latent \\
$Dec(\cdot)$ & AE decoder & Latent $\rightarrow$ image \\
$p_{\mathrm{data}}$ & Real data distribution & Dataset \\
$p_G$ & Generated data distribution & Model-induced \\
\hline
\end{tabular}
\end{center}

\section{Generic implementation of generator update iteration}
\label{appendix:pseudocode}

We present a pseudo-code showing a batch training of the GAN where both the critic and the generator are updated multiple times in a given training step.

\begin{algorithm}
\DontPrintSemicolon
\caption{Training loop for one step of the GAN}
\label{alg:gan}
\For{\textsf{\upshape train_batch} in \textsf{\upshape training_dataloader}}{
    real_lat_vec $\gets$ autoencoder.encode(train_batch)\;
    \For{\textsf{\upshape critic_iteration} in \textsf{\upshape range(n_critic)}}{
        noise $\gets$ sample_noise()\;
        fake_lat_vec $\gets$ generator(noise)\;
        critic_real_loss $\gets$ critic(real_lat_vec)\;
        critic_fake_loss $\gets$ critic(fake_lat_vec)\;
        loss_wasserstein $\gets$ mean(critic_real_loss) $-$ mean(critic_fake_loss)\;
        gradient_penalty $\gets$ calculate_gradient_penalty(critic, real_lat_vec, fake_lat_vec, lamb_gp)\;
        loss_critic $\gets$ loss_wasserstein $+$ gradient_penalty\;
        backprop(loss_critic)\;}
    \For{\textsf{\upshape generator_iteration} in \textsf{\upshape range(n_generator)}}{
        noise $\gets$ sample_noise()\;
        fake_lat_vec $\gets$ generator(noise)\;
        critic_fake_loss $\gets$ critic(fake_lat_vec)\;
        loss_generator $\gets$ $-$mean(critic_fake_loss)\;
        backprop(loss_generator)\;}}
\end{algorithm}

\pagebreak

\bibliography{apssamp}

@misc{chang2024latentstylebasedquantumgan,
      title={{Latent Style-based Quantum GAN for high-quality Image Generation}}, 
      author={Su Yeon Chang and Supanut Thanasilp and Bertrand Le Saux and Sofia Vallecorsa and Michele Grossi},
      year={2024},
      eprint={2406.02668},
      archivePrefix={arXiv},
      primaryClass={quant-ph},
      url={https://arxiv.org/abs/2406.02668}, 
}

@article{maccormack2022branching,
  title={Branching quantum convolutional neural networks},
  author={MacCormack, Ian and Delaney, Conor and Galda, Alexey and Aggarwal, Nidhi and Narang, Prineha},
  journal={Physical Review Research},
  volume={4},
  number={1},
  pages={013117},
  year={2022},
  publisher={APS},
  doi={10.1103/PhysRevResearch.4.013117},
  url={https://doi.org/10.1103/PhysRevResearch.4.013117}
}

@article{liu2020deepsat,
  title={Deepsat v2: feature augmented convolutional neural nets for satellite image classification},
  author={Liu, Qun and Basu, Saikat and Ganguly, Sangram and Mukhopadhyay, Supratik and DiBiano, Robert and Karki, Manohar and Nemani, Ramakrishna},
  journal={Remote Sensing Letters},
  volume={11},
  number={2},
  pages={156--165},
  year={2020},
  publisher={Taylor \& Francis},
  doi={10.1080/2150704X.2019.1693071},
  url={https://doi.org/10.1080/2150704X.2019.1693071}
}

@inproceedings{sat4,
author = {Basu, Saikat and Ganguly, Sangram and Mukhopadhyay, Supratik and DiBiano, Robert and Karki, Manohar and Nemani, Ramakrishna},
title = {DeepSat: a learning framework for satellite imagery},
year = {2015},
isbn = {9781450339674},
publisher = {Association for Computing Machinery},
address = {New York, NY, USA},
url = {https://doi.org/10.1145/2820783.2820816},
doi = {10.1145/2820783.2820816},
booktitle = {Proceedings of the 23rd SIGSPATIAL International Conference on Advances in Geographic Information Systems},
articleno = {37},
numpages = {10},
location = {Seattle, Washington},
series = {SIGSPATIAL '15}
}

@misc{goodfellow2014generativeadversarialnetworks,
      title={Generative Adversarial Networks}, 
      author={Ian J. Goodfellow and Jean Pouget-Abadie and Mehdi Mirza and Bing Xu and David Warde-Farley and Sherjil Ozair and Aaron Courville and Yoshua Bengio},
      year={2014},
      eprint={1406.2661},
      archivePrefix={arXiv},
      primaryClass={stat.ML},
      url={https://arxiv.org/abs/1406.2661}, 
}

@article{heusel2017gans,
  title={Gans trained by a two time-scale update rule converge to a local nash equilibrium},
  author={Heusel, Martin and Ramsauer, Hubert and Unterthiner, Thomas and Nessler, Bernhard and Hochreiter, Sepp},
  journal={Advances in neural information processing systems},
  volume={30},
  year={2017},
  url = {https://proceedings.neurips.cc/paper_files/paper/2017/file/8a1d694707eb0fefe65871369074926d-Paper.pdf},
}

@misc{Seitzer2020FID,
  author={Maximilian Seitzer},
  title={{pytorch-fid: FID Score for PyTorch}},
  month={August},
  year={2020},
  note={Version 0.3.0},
  howpublished={\url{https://github.com/mseitzer/pytorch-fid}},
}

@misc{weng2019gan,
      title={{From GAN to WGAN}}, 
      author={Lilian Weng},
      year={2019},
      eprint={1904.08994},
      archivePrefix={arXiv},
      primaryClass={cs.LG},
      url={https://arxiv.org/abs/1904.08994}, 
}

@Article{electronics14183580,
AUTHOR = {Sajjadi Mohammadabadi, Seyed Mahmoud and Kara, Burak Cem and Eyupoglu, Can and Uzay, Can and Tosun, Mehmet Serkan and Karakuş, Oktay},
TITLE = {{A Survey of Large Language Models: Evolution, Architectures, Adaptation, Benchmarking, Applications, Challenges, and Societal Implications}},
JOURNAL = {Electronics},
VOLUME = {14},
YEAR = {2025},
NUMBER = {18},
pages = {3580},
URL = {https://www.mdpi.com/2079-9292/14/18/3580},
ISSN = {2079-9292},
DOI = {10.3390/electronics14183580}
}

@article{karmakar2025role,
  title={{The role of generative AI in medical image synthesis: A review}},
  author={Karmakar, Ajay and Shaw, Abhijeet and Rakshit, Saswati and Chakraborty, Sayan and Biswas, Sitanath and Sahoo, Shubhashree and Biswas, Suparna},
  journal={Discover Applied Sciences},
  volume={7},
  number={10},
  pages={1219},
  year={2025},
  publisher={Springer},
  url = {https://doi.org/10.1007/s42452-025-07714-7},
  doi = {10.1007/s42452-025-07714-7}
}

@article{10.1016/j.cag.2023.05.010,
author = {Trevisan de Souza, Vinicius Luis and Marques, Bruno Augusto Dorta and Batagelo, Harlen Costa and Gois, Jo\~{a}o Paulo},
title = {A review on Generative Adversarial Networks for image generation},
year = {2023},
issue_date = {Aug 2023},
publisher = {Pergamon Press, Inc.},
address = {USA},
volume = {114},
number = {C},
issn = {0097-8493},
url = {https://doi.org/10.1016/j.cag.2023.05.010},
doi = {10.1016/j.cag.2023.05.010},
journal = {Comput. Graph.},
month = aug,
pages = {13–25},
numpages = {13}
}

@article{kotkondawar2025generative,
  title={A generative framework for enhancing drug target interaction prediction in drug discovery},
  author={Kotkondawar, Roshan R and Sutar, Sanjay R and Kiwelekar, Arvind W and Kadam, Vinod J and Jadhav, Shivajirao M},
  journal={Scientific Reports},
  volume={15},
  number={1},
  pages={35588},
  year={2025},
  publisher={Nature Publishing Group UK London},
  doi={10.1038/s41598-025-01589-9},
  url={https://doi.org/10.1038/s41598-025-01589-9}
}

@article{amin2018quantum,
  title={{Quantum Boltzmann machine}},
  author={Amin, Mohammad H and Andriyash, Evgeny and Rolfe, Jason and Kulchytskyy, Bohdan and Melko, Roger},
  journal={Physical Review X},
  volume={8},
  number={2},
  pages={021050},
  year={2018},
  publisher={APS},
  doi={10.1103/PhysRevX.8.021050},
  url={https://doi.org/10.1103/PhysRevX.8.021050}
}

@article{liu2018differentiable,
  title={Differentiable learning of quantum circuit born machines},
  author={Liu, Jin-Guo and Wang, Lei},
  journal={Physical Review A},
  volume={98},
  number={6},
  pages={062324},
  year={2018},
  publisher={APS},
  doi={10.1103/PhysRevA.98.062324},
  url={https://doi.org/10.1103/PhysRevA.98.062324}
}

@article{barthe2025parameterized,
  title={Parameterized quantum circuits as universal generative models for continuous multivariate distributions},
  author={Barthe, Alice and Grossi, Michele and Vallecorsa, Sofia and Tura, Jordi and Dunjko, Vedran},
  journal={npj Quantum Information},
  volume={11},
  number={1},
  pages={121},
  year={2025},
  publisher={Nature Publishing Group UK London},
  doi={10.1038/s41534-025-01064-3},
  url={https://doi.org/10.1038/s41534-025-01064-3}
}

@article{demidik2025expressive,
  title={{Expressive equivalence of classical and quantum restricted Boltzmann machines}},
  author={Demidik, Maria and T{\"u}ys{\"u}z, Cenk and Piatkowski, Nico and Grossi, Michele and Jansen, Karl},
  journal={Communications Physics},
  volume={8},
  number={1},
  pages={413},
  year={2025},
  publisher={Nature Publishing Group UK London},
  doi={10.1038/s42005-025-02353-1},
  url={https://doi.org/10.1038/s42005-025-02353-1}
}

@misc{demidik2025sample,
      title={Sample-based training of quantum generative models}, 
      author={Maria Demidik and Cenk T{\"u}ys{\"u}z and Michele Grossi and Karl Jansen},
      year={2025},
      eprint={2511.11802},
      archivePrefix={arXiv},
      primaryClass={quant-ph},
      url={https://arxiv.org/abs/2511.11802}, 
}

@article{dallaire2018quantum,
   title={Quantum generative adversarial networks},
   volume={98},
   ISSN={2469-9934},
   url={https://doi.org/10.1103/PhysRevA.98.012324},
   DOI={10.1103/physreva.98.012324},
   number={1},
   pages = {012324},
   journal={Physical Review A},
   publisher={American Physical Society (APS)},
   author={Dallaire-Demers, Pierre-Luc and Killoran, Nathan},
   year={2018},
   month=jul }

@article{lloyd2018quantum,
  title={Quantum generative adversarial learning},
  author={Lloyd, Seth and Weedbrook, Christian},
  journal={Physical review letters},
  volume={121},
  number={4},
  pages={040502},
  year={2018},
  publisher={APS},
  doi={10.1103/PhysRevLett.121.040502},
  url={https://doi.org/10.1103/PhysRevLett.121.040502}
}

@article{hu2019quantum,
  title={Quantum generative adversarial learning in a superconducting quantum circuit},
  author={Hu, Ling and Wu, Shu-Hao and Cai, Weizhou and Ma, Yuwei and Mu, Xianghao and Xu, Yuan and Wang, Haiyan and Song, Yipu and Deng, Dong-Ling and Zou, Chang-Ling and others},
  journal={Science advances},
  volume={5},
  number={1},
  pages={eaav2761},
  year={2019},
  publisher={American Association for the Advancement of Science},
  url={https://doi.org/10.1126/sciadv.aav2761},
  doi={10.1126/sciadv.aav2761}
}

@article{zoufal2019quantum,
  title={Quantum generative adversarial networks for learning and loading random distributions},
  author={Zoufal, Christa and Lucchi, Aur{\'e}lien and Woerner, Stefan},
  journal={npj Quantum Information},
  volume={5},
  number={1},
  pages={103},
  year={2019},
  publisher={Nature Publishing Group UK London},
  doi={doi.org/10.1038/s41534-019-0223-2},
  url={https://doi.org/10.1038/s41534-019-0223-2}
}

@article{niu2022entangling,
  title={Entangling quantum generative adversarial networks},
  author={Niu, Murphy Yuezhen and Zlokapa, Alexander and Broughton, Michael and Boixo, Sergio and Mohseni, Masoud and Smelyanskyi, Vadim and Neven, Hartmut},
  journal={Physical review letters},
  volume={128},
  number={22},
  pages={220505},
  year={2022},
  publisher={APS},
  doi={10.1103/PhysRevLett.128.220505},
  url={https://doi.org/10.1103/PhysRevLett.128.220505}
}

@article{chaudhary2023towards,
  title={Towards a scalable discrete quantum generative adversarial neural network},
  author={Chaudhary, Smit and Huembeli, Patrick and MacCormack, Ian and Patti, Taylor L and Kossaifi, Jean and Galda, Alexey},
  journal={Quantum Science and Technology},
  volume={8},
  number={3},
  pages={035002},
  year={2023},
  publisher={IOP Publishing},
  doi={10.1088/2058-9565/acc4e4},
  url={https://doi.org/10.1088/2058-9565/acc4e4}
}

@article{Diptanshu_Sikdar,
  title={{Hybrid Quantum-Classical Generative Adversarial Network for synthesizing chemically feasible molecules}},
  author={Diptanshu Sikdar and Max Cui and Adelina Chau and Arjun Bhamra and Sathvik Prasanna and Larry McMahan},
  journal={Journal of emerging investigations},
  volume={6},
  number={6},
  pages={1},
  year={2023},
  doi={10.59720/22-143},
  ur={https://doi.org/10.59720/22-143}
}

@ARTICLE{9520764,
  author={Li, Junde and Topaloglu, Rasit O. and Ghosh, Swaroop},
  journal={IEEE Transactions on Quantum Engineering}, 
  title={{Quantum Generative Models for Small Molecule Drug Discovery}}, 
  year={2021},
  volume={2},
  number={},
  pages={1-8},
  doi={10.1109/TQE.2021.3104804},
  url={https://doi.org/10.1109/TQE.2021.3104804}}

@ARTICLE{10556803,
  author={Anoshin, Matvei and Sagingalieva, Asel and Mansell, Christopher and Zhiganov, Dmitry and Shete, Vishal and Pflitsch, Markus and Melnikov, Alexey},
  journal={IEEE Transactions on Quantum Engineering}, 
  title={{Hybrid Quantum Cycle Generative Adversarial Network for Small Molecule Generation}}, 
  year={2024},
  volume={5},
  number={},
  pages={1-14},
  doi={10.1109/TQE.2024.3414264},
  url={https://doi.org/10.1109/TQE.2024.3414264}}

@article{mhiri2025constrained,
  title={{Constrained and Vanishing Expressivity of Quantum Fourier Models}},
  author={Mhiri, Hela and Monbroussou, Leo and Herrero-Gonzalez, Mario and Thabet, Slimane and Kashefi, Elham and Landman, Jonas},
  journal={Quantum},
  volume={9},
  pages={1847},
  year={2025},
  publisher={Verein zur F{\"o}rderung des Open Access Publizierens in den Quantenwissenschaften},
  url={https://doi.org/10.22331/q-2025-09-03-1847},
  DOI={10.22331/q-2025-09-03-1847},
}

@article{jaderberg2024let,
  title={Let quantum neural networks choose their own frequencies},
  author={Jaderberg, Ben and Gentile, Antonio A and Berrada, Youssef Achari and Shishenina, Elvira and Elfving, Vincent E},
  journal={Physical Review A},
  volume={109},
  number={4},
  pages={042421},
  year={2024},
  publisher={APS},
  doi={10.1103/PhysRevA.109.042421},
  url={https://doi.org/10.1103/PhysRevA.109.042421}
}

@misc{xu2024frequency,
      title={{Frequency principle for quantum machine learning via Fourier analysis}}, 
      author={Yi-Hang Xu and Dan-Bo Zhang},
      year={2025},
      eprint={2409.06682},
      archivePrefix={arXiv},
      primaryClass={quant-ph},
      url={https://arxiv.org/abs/2409.06682}, 
}

@misc{duffy2025spectral,
      title={{Spectral Bias in Variational Quantum Machine Learning}}, 
      author={Callum Duffy and Marcin Jastrzebski},
      year={2025},
      eprint={2506.22555},
      archivePrefix={arXiv},
      primaryClass={quant-ph},
      url={https://arxiv.org/abs/2506.22555}, 
}

@misc{bermejo2024quantumconvolutionalneuralnetworks,
      title={{Quantum Convolutional Neural Networks are (Effectively) Classically Simulable}}, 
      author={Pablo Bermejo and Paolo Braccia and Manuel S. Rudolph and Zoë Holmes and Lukasz Cincio and M. Cerezo},
      year={2024},
      eprint={2408.12739},
      archivePrefix={arXiv},
      primaryClass={quant-ph},
      url={https://arxiv.org/abs/2408.12739}, 
}

@article{schuld2019quantumfeature,
  title = {{Quantum Machine Learning in Feature Hilbert Spaces}},
  author = {Schuld, Maria and Killoran, Nathan},
  journal = {Physical Review Letters},
  volume = {122},
  number = {4},
  pages = {040504},
  year = {2019},
  doi = {10.1103/PhysRevLett.122.040504},
  url = {https://doi.org/10.1103/PhysRevLett.122.040504}
}

@article{gil2024understanding,
   title={Understanding quantum machine learning also requires rethinking generalization},
   volume={15},
   ISSN={2041-1723},
   url={https://doi.org/10.1038/s41467-024-45882-z},
   DOI={10.1038/s41467-024-45882-z},
   number={1},
   pages={2277},
   journal={Nature Communications},
   publisher={Springer Science and Business Media LLC},
   author={Gil-Fuster, Elies and Eisert, Jens and Bravo-Prieto, Carlos},
   year={2024},
   month=mar }

@article{Haug2024,
   title={{Generalization of Quantum Machine Learning Models Using Quantum Fisher Information Metric}},
   volume={133},
   ISSN={1079-7114},
   url={https://doi.org/10.1103/PhysRevLett.133.050603},
   DOI={10.1103/physrevlett.133.050603},
   number={5},
   pages = {050603},
   journal={Physical Review Letters},
   publisher={American Physical Society (APS)},
   author={Haug, Tobias and Kim, M. S.},
   year={2024},
   month=jul }

@inproceedings{gulrajani2017improvedtrainingwassersteingans,
 author = {Gulrajani, Ishaan and Ahmed, Faruk and Arjovsky, Martin and Dumoulin, Vincent and Courville, Aaron C},
 booktitle = {Advances in Neural Information Processing Systems},
 editor = {I. Guyon and U. Von Luxburg and S. Bengio and H. Wallach and R. Fergus and S. Vishwanathan and R. Garnett},
 pages = {},
 publisher = {Curran Associates, Inc.},
 title = {{Improved Training of Wasserstein GANs}},
 url = {https://proceedings.neurips.cc/paper_files/paper/2017/file/892c3b1c6dccd52936e27cbd0ff683d6-Paper.pdf},
 volume = {30},
 year = {2017}
}

@article{Tang2024,
    author = {Tang, Xiangru and Dai, Howard and Knight, Elizabeth and Wu, Fang and Li, Yunyang and Li, Tianxiao and Gerstein, Mark},
    title = {{A survey of generative {AI} for de novo drug design: new frontiers in molecule and protein generation}},
    journal = {Briefings in Bioinformatics},
    volume = {25},
    number = {4},
    pages = {bbae338},
    year = {2024},
    month = {07},
    issn = {1477-4054},
    doi = {10.1093/bib/bbae338},
    url = {https://doi.org/10.1093/bib/bbae338}
}

@ARTICLE{Gangwal2024,  
AUTHOR={Gangwal, Amit  and Ansari, Azim  and Ahmad, Iqrar  and Azad, Abul Kalam  and Kumarasamy, Vinoth  and Subramaniyan, Vetriselvan  and Wong, Ling Shing },
TITLE={{Generative artificial intelligence in drug discovery: basic framework, recent advances, challenges, and opportunities}},
JOURNAL={Frontiers in Pharmacology},
VOLUME={15},
YEAR={2024},
pages={1},
URL={https://www.frontiersin.org/journals/pharmacology/articles/10.3389/fphar.2024.1331062},
DOI={10.3389/fphar.2024.1331062},
ISSN={1663-9812},
}

@article{TRIPATHI2022100045,
title = {Recent advances and application of generative adversarial networks in drug discovery, development, and targeting},
journal = {Artificial Intelligence in the Life Sciences},
volume = {2},
pages = {100045},
year = {2022},
issn = {2667-3185},
doi = {https://doi.org/10.1016/j.ailsci.2022.100045},
url = {https://www.sciencedirect.com/science/article/pii/S2667318522000150},
author = {Satvik Tripathi and Alisha Isabelle Augustin and Adam Dunlop and Rithvik Sukumaran and Suhani Dheer and Alex Zavalny and Owen Haslam and Thomas Austin and Jacob Donchez and Pushpendra Kumar Tripathi and Edward Kim},
}

@Article{Saad2024,
author={Saad, Muhammad Muneeb
and O'Reilly, Ruairi
and Rehmani, Mubashir Husain},
title={A survey on training challenges in generative adversarial networks for biomedical image analysis},
journal={Artificial Intelligence Review},
year={2024},
month={Jan},
day={29},
volume={57},
number={2},
pages={19},
issn={1573-7462},
doi={10.1007/s10462-023-10624-y},
url={https://doi.org/10.1007/s10462-023-10624-y}
}

@InProceedings{Arjovsky2017,
  title = 	 {{{W}asserstein Generative Adversarial Networks}},
  author =       {Martin Arjovsky and Soumith Chintala and L{\'e}on Bottou},
  booktitle = 	 {Proceedings of the 34th International Conference on Machine Learning},
  pages = 	 {214--223},
  year = 	 {2017},
  editor = 	 {Precup, Doina and Teh, Yee Whye},
  volume = 	 {70},
  series = 	 {Proceedings of Machine Learning Research},
  month = 	 {06--11 Aug},
  publisher =    {PMLR},
  url = 	 {https://proceedings.mlr.press/v70/arjovsky17a.html}
}

@Article{Prykhodko2019,
author={Prykhodko, Oleksii
and Johansson, Simon Viet
and Kotsias, Panagiotis-Christos
and Ar{\'u}s-Pous, Josep
and Bjerrum, Esben Jannik
and Engkvist, Ola
and Chen, Hongming},
title={A de novo molecular generation method using latent vector based generative adversarial network},
journal={Journal of Cheminformatics},
year={2019},
month={Dec},
day={03},
volume={11},
number={1},
pages={74},
issn={1758-2946},
doi={10.1186/s13321-019-0397-9},
url={https://doi.org/10.1186/s13321-019-0397-9}
}

@ARTICLE{8977347,
  author={Karras, Tero and Laine, Samuli and Aila, Timo},
  journal={IEEE Transactions on Pattern Analysis and Machine Intelligence}, 
  title={{A Style-Based Generator Architecture for Generative Adversarial Networks}}, 
  year={2021},
  volume={43},
  number={12},
  pages={4217-4228},
  doi={10.1109/TPAMI.2020.2970919},
  url={https://doi.org/10.1109/TPAMI.2020.2970919}
}

@INPROCEEDINGS{9156570,
  author={Karras, Tero and Laine, Samuli and Aittala, Miika and Hellsten, Janne and Lehtinen, Jaakko and Aila, Timo},
  booktitle={2020 IEEE/CVF Conference on Computer Vision and Pattern Recognition (CVPR)}, 
  title={{Analyzing and Improving the Image Quality of StyleGAN}}, 
  year={2020},
  volume={},
  number={},
  pages={8107-8116},
  doi={10.1109/CVPR42600.2020.00813},
  url={https://doi.org/10.1109/CVPR42600.2020.00813}}

@article{BravoPrieto2022,
   title={{Style-based quantum generative adversarial networks for Monte Carlo events}},
   volume={6},
   ISSN={2521-327X},
   url={https://doi.org/10.22331/q-2022-08-17-777},
   DOI={10.22331/q-2022-08-17-777},
   journal={Quantum},
   publisher={Verein zur Forderung des Open Access Publizierens in den Quantenwissenschaften},
   author={Bravo-Prieto, Carlos and Baglio, Julien and Cè, Marco and Francis, Anthony and Grabowska, Dorota M. and Carrazza, Stefano},
   year={2022},
   month=aug, pages={777} }

@misc{baglio2024,
      title={Data augmentation experiments with style-based quantum generative adversarial networks on trapped-ion and superconducting-qubit technologies}, 
      author={Julien Baglio},
      year={2024},
      eprint={2405.04401},
      archivePrefix={arXiv},
      primaryClass={quant-ph},
      url={https://arxiv.org/abs/2405.04401}, 
}

@booklet{baglio2025,
    title={{Latent-Style-Based Generative Quantum Model for Assisted Drug Discovery}},
    author={Baglio, Julien and Haddad, Yacine and Polifka, Richard},
    year={2025},
    howpublished={Research Highlights at the Applied Quantum Methods for Computer
                  Science and Engineering (AQMCSE) Conference, Aachen, Germany, 8-10 October 2025}
}

@inproceedings{NIPS2016_8a3363ab,
 author = {Salimans, Tim and Goodfellow, Ian and Zaremba, Wojciech and Cheung, Vicki and Radford, Alec and Chen, Xi and Chen, Xi},
 booktitle = {Advances in Neural Information Processing Systems},
 editor = {D. Lee and M. Sugiyama and U. Luxburg and I. Guyon and R. Garnett},
 pages = {},
 publisher = {Curran Associates, Inc.},
 title = {Improved Techniques for Training GANs},
 url = {https://proceedings.neurips.cc/paper_files/paper/2016/file/8a3363abe792db2d8761d6403605aeb7-Paper.pdf},
 volume = {29},
 year = {2016}
}

@ARTICLE{jsd1991,
  author={Lin, J.},
  journal={IEEE Transactions on Information Theory}, 
  title={{Divergence measures based on the Shannon entropy}}, 
  year={1991},
  volume={37},
  number={1},
  pages={145-151},
  doi={10.1109/18.61115},
  url={https://doi.org/10.1109/18.61115}
}

@inproceedings{kingma2015,
  author={Kingma, Diederik P.  and Ba, Jimmy},
  editor = {Yoshua Bengio and Yann LeCun},
  title = {{Adam: A Method for Stochastic Optimization}},
  booktitle = {Conference Track Proceedings  of the 3rd International Conference on Learning Representations, {ICLR} 2015},
  address = {San Diego, CA, USA},
  month = {7--9 May},
  year = {2015},
  url = {http://arxiv.org/abs/1412.6980},
  eprint={1412.6980},
  archivePrefix={arXiv},
  primaryClass={cs.LG},
}

@article{vandenBerg2022,
   title={Model-free readout-error mitigation for quantum expectation values},
   volume={105},
   ISSN={2469-9934},
   url={http://dx.doi.org/10.1103/PhysRevA.105.032620},
   DOI={10.1103/physreva.105.032620},
   number={3},
   journal={Physical Review A},
   publisher={American Physical Society (APS)},
   author={van den Berg, Ewout and Minev, Zlatko K. and Temme, Kristan},
   year={2022},
   month=mar }

@article{Temme2017,
   title={Error Mitigation for Short-Depth Quantum Circuits},
   volume={119},
   ISSN={1079-7114},
   url={http://dx.doi.org/10.1103/PhysRevLett.119.180509},
   DOI={10.1103/physrevlett.119.180509},
   number={18},
   journal={Physical Review Letters},
   publisher={American Physical Society (APS)},
   author={Temme, Kristan and Bravyi, Sergey and Gambetta, Jay M.},
   year={2017},
   month=nov }

\end{document}